\documentclass[12pt]{article}

\usepackage[english]{babel}
\usepackage[latin1]{inputenc}
\usepackage{lmodern,csquotes}
\usepackage[T1]{fontenc}

\usepackage[a4paper,left=0.7in,right=0.7in,top=1in,bottom=1in]{geometry}

\usepackage[backend=biber,style=authoryear-comp,maxcitenames=1,maxbibnames=99,sorting=nyt,sortcites=false]{biblatex}
\usepackage{amssymb, amsmath, amsthm, mathtools, dsfont, xfrac, bm}
\usepackage{graphicx, tikz, pgfplots, pgfplotstable, booktabs, multirow, enumitem}
\pgfplotsset{compat=newest}
\usepgfplotslibrary{fillbetween}

\usepackage[section]{placeins}

\usepgfplotslibrary{fillbetween}
\pgfplotsset{
  compat=newest,
  layers/axis lines on top/.define layer set={
    axis background,
    axis grid,
    axis ticks,
    axis tick labels,
    pre main,
    main,
    axis lines,
    axis descriptions,
    axis foreground,
  }{/pgfplots/layers/standard},
}

\pgfkeys{
    /pgfplots/table/print last/.style={
        row predicate/.code={
            \pgfmathtruncatemacro\firstprintedrownumber{\pgfplotstablerows-#1} 
            \ifnum##1<\firstprintedrownumber\relax
                \pgfplotstableuserowfalse
            \fi
        }
    },
    /pgfplots/table/print last/.default=1
}
\pgfplotsset{select coords between index/.style 2 args={
    x filter/.code={
        \ifnum\coordindex<#1\fi
        \ifnum\coordindex>#2\fi
    }
}}

\pgfplotstableread[col sep=semicolon, trim cells]{asian_floating_strike.dat}\afsdat
\pgfplotstableread[col sep=semicolon, trim cells]{asian_fixed_strike.dat}\axsdat
\pgfplotstableread[col sep=semicolon, trim cells]{lookback_floating_strike.dat}\lfsdat
\pgfplotstableread[col sep=semicolon, trim cells]{lookback_fixed_strike.dat}\lxsdat
\pgfplotstableread[col sep=semicolon, trim cells]{certificate.dat}\certificatedat
\pgfplotstableread{american-greeks-mc.dat}\sensmc
\pgfplotstableread{asian-greeks-mc.dat}\sensmcaxs
\pgfplotstableread{snowball-greeks-mc.dat}\sensmcsnw
\pgfplotstableread{lockin-greeks-mc.dat}\sensmclck
\pgfplotstableread{sensitivity_american_cmp_2.dat}\senscmp
\pgfplotstableread{asian_fixed_strike_sensitivities.dat}\sensaxs
\pgfplotstableread{lookback_fixed_strike_sensitivity.dat}\senslxs
\pgfplotstableread{amerasian_w2.dat}\sensaxscmp
\pgfplotstableread{snowball_w2.dat}\senssnwcmp
\pgfplotstableread{lockin_w2.dat}\senslckcmp

\definecolor{mylightyellow}{rgb}{1,1,.8}
\definecolor{mylightgreen}{rgb}{.8,1,.8}
\definecolor{mydarkorange}{RGB}{217,126,29}
\definecolor{mydarkred}{RGB}{178,34,34}
\definecolor{mydarkgreen}{RGB}{34,139,34}
\definecolor{mydarkblue}{RGB}{72,61,139}
\definecolor{mydarkyellow}{RGB}{218,165,32}

\usepackage[normalem]{ulem}
\newcommand\soutpars[1]{\let\helpcmd\sout\parhelp#1\par\relax\relax}
\long\def\parhelp#1\par#2\relax{%
  \helpcmd{#1}\ifx\relax#2\else\par\parhelp#2\relax\fi%
}
\usepackage{hyperref}
\hypersetup{
 colorlinks=true,breaklinks=true,
 urlcolor= mydarkblue,linkcolor= mydarkblue,citecolor= mydarkgreen,
 pdfauthor={Gambara, M., Livieri, G., Pallavicini, A.}
}

\theoremstyle{plain}
\newtheorem{theorem}{Theorem}[section]

\newtheorem{lemma}[theorem]{Lemma}
\newtheorem{proposition}[theorem]{Proposition}
\theoremstyle{definition}
\newtheorem{definition}{Definition}
\theoremstyle{remark}

\DeclarePairedDelimiterX\QCond[2]{\{}{\}}{\,#1\,\delimsize\vert\,\mathopen{}#2\,}
\DeclarePairedDelimiterX\QUncond[1]{\{}{\}}{\,#1\,}
\DeclarePairedDelimiterX\ECond[2]{[}{]}{\,#1\,\delimsize\vert\,\mathopen{}#2\,}
\DeclarePairedDelimiterX\EUncond[1]{[}{]}{\,#1\,}

\newcommand{\ExO}[1]{\mathds{E}\EUncond*{#1}}

\newcommand{\ExF}[2]{\mathds{E}\ECond*{#2}{{\cal F}_{#1}}}

\newcommand{\std}[1]{{\rm Std}\EUncond*{#1}}
\newcommand{\ind}[1]{\mathds{1}_{\{#1\}}}
\newcommand{\risk}[1]{R^{#1}}

\begin{document}

\title{Machine-learning regression methods\\ for American-style path-dependent contracts\footnote{The authors report no potential competing interests. The opinions expressed in this document are solely those of the authors and do not represent in any way those of their present and past employers.}}

\author{M.~Gambara\thanks{Inait SA, Address: Av. du Tribunal-F\'ed\'eral 34, 1005, Lausanne, Switzerland. Email address: \texttt{matteo.gambara@gmail.com}.}, 
G.~Livieri\thanks{The London School of Economics and Political Science, Department of Statistics. Address: Houghton St, London WC2A 2AE, United Kingdom. Email address: \texttt{g.livieri@lse.ac.uk}.}, 
A.~Pallavicini\thanks{Intesa Sanpaolo, Financial Engineering. Address: largo Mattioli 3, Milano 20121, Italy. Email address: \texttt{andrea.pallavicini@intesasanpaolo.com}.}}

\date{
\small First Version: January 31, 2023.  This version: \today
}

\maketitle

\begin{abstract}
  Evaluating financial products with early-termination clauses, in particular those with path-dependent structures, is challenging. This paper focuses on Asian options, look-back options, and callable certificates. We will compare regression methods for pricing and computing sensitivities, highlighting modern machine learning techniques against traditional polynomial basis functions. Specifically, we will analyze randomized recurrent and feed-forward neural networks, along with a novel approach using signatures of the underlying price process. For option sensitivities like Delta and Gamma, we will incorporate Chebyshev interpolation. Our findings show that machine learning algorithms often match the accuracy and efficiency of traditional methods for Asian and look-back options, while randomized neural networks are best for callable certificates. Furthermore, we apply Chebyshev interpolation for Delta and Gamma calculations for the first time in Asian options and callable certificates.
\end{abstract}

\bigskip

\noindent {\bf AMS classification codes:} 65C05, 91G20, 91G60.

\newpage
{\small \tableofcontents}
\vfill
\newpage
{\small \listoftables \listoffigures}
\vfill
\newpage

\pagestyle{myheadings} \markboth{}{{\footnotesize  Gambara, Livieri, Pallavicini, American-style path-dependent contracts}}

\section{Introduction}
\label{sec:intro}

Financial products with early-termination features have become increasingly popular. These products allow investors to participate in the performance of underlying assets while granting them the flexibility to exit their positions before the maturity date. The investor or the issuer may have the right to terminate the contract, subject to specific conditions. These products are commonly structured as derivatives contracts or certificates. The early-termination options within these products allow for swift liquidation triggered by factors such as asset performance or specific events.

Evaluating and calculating sensitivities for financial products with early-termination clauses presents significant challenges, especially when these products have path-de\-pen\-dent structures. Traditional techniques, such as least-squares Monte Carlo (LSMC) simulations, can struggle with the ``curse of dimensionality." LSMC relies on ordinary least-squares (OLS) approximation, which requires the selection of appropriate \textit{basis functions}. As the dimensionality of the regression basis increases, the performance of LSMC deteriorates, growing polynomially or even exponentially with the number of risk factors, as discussed in Section~2.2 of~\textcite{Longstaff2001}. Therefore, the selection of the basis function is crucial for the analysis in this paper.

The aim of this paper is to examine the performance of various regression methods in pricing and computing sensitivities. We will discuss the advantages and limitations of modern approaches made possible by recent advances in machine learning, which serve as alternatives to traditional polynomial basis functions. Specifically, we will enhance regression-based methods by introducing alternative basis function choices derived from randomized neural networks and signatures, and we will analyze whether this approach offers a promising solution to the challenges posed by path-dependent early-terminating financial products such as Asian options, look-back options, and callable certificates. Instead, for computing option sensitivities, particularly Delta and Gamma, this paper complements these techniques with Chebyshev interpolation.

Asian and look-back options are usually traded in over-the-counter markets, providing investors with opportunities to capitalize on the average or extreme values of underlying asset prices over a predetermined period. The time window associated with these options can either be fixed, where the average or extreme value is calculated over a specific period, or rolling, where the calculation period continuously updates as time progresses. In certain cases, these contracts may also incorporate early-termination features, allowing investors to exit the position before expiration, as described in~\textcite{Kao2003, Bernard2014, Goudenege2022}. In the context of commodity markets, these types of options can be regarded as exemplifying Swing contracts. Swing contracts allow the holder to adjust the quantity of the underlying asset delivered or received within a specified time frame. We refer to~\textcite{Thompson1995,Barrera2006,Daluiso2023} for further details.

The dimensionality issue, mentioned above, is especially pronounced in the presence of path dependence. Alternative approaches explored in the literature include partial differential equation methods (e.g., \textcite{dai2010lattice,federico2015finite}) and lattice-based techniques (e.g., \textcite{bernhart2011finite,lelong2019pricing}), with the latter often applied to price options on averages. However, such methods are typically constrained to settings with limited complexity, for instance, moving-average options with a monitoring window of at most ten observations (e.g., \textcite{bernhart2011finite,lelong2019pricing}).

Recently, the problem of pricing early-terminating products via LSMC has been addressed by replacing the traditional basis functions with neural networks and employing gradient descent instead of OLS; see, for example, \textcite{Kohler2010, Becker2020, Lapeyre2021}. However, since neural networks are non-convex functions with respect to their trainable parameters, gradient descent does not necessarily converge to a global minimum under typical training procedures (e.g., finite training time, explicit regularization). In contrast, OLS minimization of linear systems guarantees global convergence. This lack of convergence guarantees is a significant drawback of these methods, as they often rely on strong and unrealistic assumptions to establish theoretical results. Additionally, these approaches necessitate an off-line training phase that must be repeated whenever market conditions change or a different payoff structure is considered, further complicating their practical implementation.

A tentative to overcome these problems is proposed in~\textcite{herrera2021optimal}, where randomized neural networks are employed as a linear regression basis. A randomized neural network can be viewed as a linear combination of random basis functions. In this approach (e.g., \textcite{cao2018review}), instead of training all layers of the neural network, the parameters of the hidden layers are randomly initialized and kept fixed, while only the parameters of the output layer are optimized. This significantly accelerates the training process, enabling it to be performed on-line whenever a price is required.

In the present paper, as a first contribution, we adapt the approach of~\textcite{herrera2021optimal} to deal with the previously discussed payoffs, and we compare the results with an alternative original formulation based on signature methods. In particular, we employ both non-randomized signatures and randomized signatures. The former are novel mathematical tools that have been developed to study complex and irregular paths and functions. They represent a path in terms of its iterated integrals, hence capturing essential information about path roughness and regularity. In the present context, non-randomized signature methods offer a unique advantage as they provide a way to express the state path dependency in terms of a linear function of a limited number of risk factors so that we can apply standard regression techniques. However, they may suffer from similar disadvantages as polynomials, such as the ``curse of dimensionality'', which can be overcome by randomized signatures, i.e.~algorithms that try to approximate input/output systems without the need of a full calibration of the system itself.

The literature on non-randomized signature methods with applications in finance is huge, so that, in this contribution, we limit ourselves to cite only the most recent research papers on the topic focusing, to the best of our knowledge, on the pricing of financial derivatives. \textcite{sabate2020solving} use a combination of recurrent neural network and signature methods to design efficient algorithms for solving parametric families of path-dependent partial differential equations (PDE) that arise in pricing and hedging of path-dependent derivatives or from use of non-Markov models. A path-dependent PDE solver, based on signature kernels, is also proposed in~\textcite{pannier2024path}. On the other hand, the pricing of exotic vanilla derivatives is covered in~\textcite{lyons2019numerical}. Finally, methods based on signatures for pricing path-dependent options can be found also in~\textcite{Feng2021sig} and~\textcite{Bayraktar2022}, where signatures are used as features for either feed-forward or recurrent neural networks. We claim that the use of deep neural networks here is not strictly necessary, whence the novelty of our approach when using signatures as basis functions, because of the ability of the signature by itself to represent data using a small set of features, by capturing the most important properties of data in a non-parametric way. On the other hand, examples of applications of randomized signatures can be found in~\textcite{Akyildirim2022anomaly}, where they are used to identify pump-and-dump attempts in the crypto market in an unsupervised learning settings, and in~\textcite{Akyildirim2023portfolio}, where randomized signatures are utilized as a non-parametric and non-linear drift estimator to find an optimal allocation a long-only portfolio. Non-randomized and randomized signature methods, along with the corresponding relevant mathematical literature, are discusses and reported in Section~\ref{sec:signature}

As a second contribution, we then analyze algorithms to compute sensitivities, particularly Delta and Gamma. As shown in~\textcite{herrera2021optimal}, second-order sensitivities are quite noisy when regression methods are used for pricing, and they require particular care to be calculated in a robust way. We investigate the regression-based method by~\textcite{Letourneau2019} and the Chebyshev interpolation techniques developed in~\textcite{Maran2020}. To the best of our knowledge, it is the first time that this analysis is performed for American-style Asian and look-back options and for callable certificates.

Our numerical results indicate that machine learning-based algorithms achieve accuracy and computational efficiency that are comparable to traditional algorithms when pricing Asian and look-back options. In contrast, we find that randomized neural networks are the most effective choice of basis functions for pricing callable certificates. Finally, we discovered that Chebyshev interpolation techniques can be successfully applied, for the first time, to Delta and Gamma calculations for both Asian options and callable certificates.

\medskip

The paper is organized as follows. In Section~\ref{sec:payoff} we discuss the payoffs under investigation, namely American-style Asian and look-back options. In Section~\ref{sec:pricing}, we present the pricing algorithms. We start by introducing the standard approach based on the LSMC; then we continue with the randomized neural networks, and the signature methods. We end the section by discussing how to compute sensitivities. In Section~\ref{sec:results} we present the numerical details of the algorithms and we perform investigations on the different payoffs we have presented. As a last contribution we discuss the case of callable certificates in Section~\ref{sec:callable}. Then, in Section~\ref{sec:conclusion} we summarize the paper and we hint at future developments. The paper concludes with different appendices that should deepen some theoretical aspects of the employed algorithms (Appendix~\ref{app:Zanette} and~\ref{app:signature}), the theory of other derivatives with similar features, such as Snowball and Lock-in callable certificates (Appendix~\ref{app:callable}), the numerical methods of sensitivity calculations (Appendix~\ref{app:sensitivity}), and provide more numerical results on the proposed methods for the interested reader (Appendix~\ref{app:numerical_data}).

\subsection{Notation}
\label{sec:notation}

For the sake of the reader, we collect and define here \textit{some} notations that we will use in the rest of the paper, and we indicate the exact point where the first appearance of a symbol occurs:
\begin{enumerate}
\item[1.] $M$: the length of the moving window over which the average, minimum or maximum of the price process is computed when pricing Asian and look-back payoffs; \ref{sec:amerasian}. 
\item[2.] $\{T_0,\ldots,T_N\}$, where $T_0:=0$ and $T_N:=T$: observation dates; \ref{sec:amerasian}. 
\item[3.] $V_{T_i}$: value function at $T_i$; Equation \eqref{eq:asian_dynprog}; $C_{T_i}$: continuation value at $T_i$; Equation \eqref{eq:continuation}. $B_{T_i}$: bank-account process at $T_i$; Equation \eqref{eq:asian_dynprog}. 
\item[4.] $\{\varphi_1,\ldots,\varphi_B\}$: set of $B$ basis functions; just below Equation \eqref{eq:continuation_approx}.
\item[5.] $h$: hidden size of random neural networks; Subsection~\ref{sec:rnn}.
\item[6.] $d$: dimension of the state process at each observation date $T_i$; Subsections~\ref{sec:rnn} and~\ref{sec:signature}, and also Appendix~\ref{app:signature}.
\item[7.] $n$: order of the truncation of the signature; Subsection~\ref{sec:signature} and Appendix~\ref{app:signature}.
\item[8.] $k^{\star}$: dimension of the reservoir system in the randomized signature approach; Equation \eqref{eq:ran_sig}.
\item[9.] $\varsigma$: variance of the independent normal random variables used to populate the random matrices in the randomized signature approach; Subsection~\ref{sec:rand_signature}. 
\item[10.] $P$ and $Q$: number of discrete paths for the price process via Gaussian Process Regression and number of points employed in the Gauss Hermite Quadrature; Subsection~\ref{sec:asian_options}.
\item[11.] $R^{\rho}$ with $\rho \in \{1,2,3,4\}$: sets of risk factors; \ref{sec:asian_options}.
\end{enumerate}

\section{Financial products with early termination}
\label{sec:payoff}

In the present paper, we discuss numerical algorithms to price path-dependent financial products with early-termination clauses. In this section, we describe the selection of payoffs considered in our main analysis: Asian and look-back options.

\subsection{Asian and look-back payoffs}
\label{sec:amerasian}

In many markets, ranging from equity to commodity exchanges, other-the-counter options are typically sold with early-termination clauses. Call and put options, as well as Asian and look-back options may share such clauses.

In general, we consider options up to a maturity date $T$ written on an underlying asset, whose price process we term $S_t$ with $t\in[0,T]$. These options are written either on the average, minimum or maximum of the price process over a moving window of $M$ days. In order to write their payoff, we first introduce a grid of daily observation dates $\{T_0, \ldots, T_N\}$, where $T_0:=0$ and $T_N:=T$, and the following random variables
\begin{equation}
\label{eq:movingaverage}
 A^{\rm avg}_{T_i}(M) := \frac{1}{M} \sum_{j=i-M+1}^i S_{T_j},
\end{equation}%
along with
\begin{equation}
 A^{\rm min}_{T_i}(M) := \min_{j\in[i-M+1,i]} S_{T_j}, \quad
 A^{\rm max}_{T_i}(M) := \max_{j\in[i-M+1,i]} S_{T_j},
\end{equation}%
defined for $M-1 \le i \le N$.

Then, we can define the option payoff we wish to discuss. We start with the fixed-strike and floating-strike put Asian payoffs, respectively given by
\begin{equation}
\label{eq:asian}
 \Psi^{\rm A1}_{T_i}(M;K) := \max \left(K - A^{\rm avg}_{T_i}(M), 0 \right), \quad
 \Psi^{\rm A2}_{T_i}(M) := \max \left(S_{T_i} - A^{\rm avg}_{T_i}(M), 0 \right),
\end{equation}%
where $K>0$ is the strike price. Moreover, we introduce the corresponding look-back payoffs, given by
\begin{equation}
\label{eq:lookback}
 \Psi^{\rm L1}_{T_i}(M;K) := \max \left(K - A^{\rm max}_{T_i}(M), 0 \right), \quad
 \Psi^{\rm L2}_{T_i}(M) := \max \left(S_{T_i} - A^{\rm min}_{T_i}(M), 0 \right).
\end{equation}%

In the following, we will use the symbol $\Psi_{T_i}(M)$ to denote them if it is clear from the context to which payoff we are referring. We observe that since we consider that the first available underlying value is $S_0$, the payoff function cannot be evaluated before time $T_{M-1}$. Our option can be exercised at any time instant $T_i$, for $i = M,\ldots,N$. It is worth noticing that, following~\textcite{bernhart2011finite}, \textcite{lelong2019pricing} and~\textcite{Goudenege2022}, the first time the option can be exercises is $T_M$, but other choices are possible. In the above definitions we follow~\textcite{bernhart2011finite} and we always assume that $M-1 < i \le N$, so that we never consider payoffs with moving windows in the past.

\subsection{Early exercise and early termination}
\label{sec:early}

The payoffs described in the previous subsection are paid by financial products which can be terminated before the contract maturity date $T$ by the option buyer.

In this section, we setup the pricing framework for this products. We choose a risk-neutral pricing model $(\Omega,\mathcal{F},\mathds{Q})$, with a filtration $(\mathcal{F}_t)_{t \in [0,T]}$ representing all the observable market quantities. Thus, the price processes $S_t$ are adapted to such a filtration.

We consider that the options can be exercised on any date in the time grid $\{T_M, \ldots,$ $T_N\}$ previously defined. We setup the pricing problem by starting from the last exercise date $T_N$, and we introduce the ${\cal F}_{T_N}$-measurable value function $V_N$, defined as
\begin{equation}
 V_{T_N} := \Psi_{T_N} ,
\end{equation}%
where $\Psi$ is one of the Asian or look-back payoffs introduced in the previous Subsection~\ref{sec:amerasian} (see Equations~\eqref{eq:asian} and~\eqref{eq:lookback}); here we have not made explicit the dependence on the parameters $M$ and $K$ in $\Psi$, for the ease of notation. On the previous dates $T_i$, with $M-1<i<N$ the option buyer may decide to exercise the option, by choosing the maximum between the immediate exercise and the continuation value of the contract, precisely defined below. Thus, we can write
\begin{equation}
\label{eq:asian_dynprog}
 V_{T_i} := \max \left\{ \Psi_{T_i}, B_{T_i} \,\ExF{T_i}{ \frac{V_{T_{i+1}}}{B_{T_{i+1}}} } \right\},
\end{equation}%
where $B_t$ is the bank-account process, and the expectation is taken under the pricing measure $\mathds{Q}$. We assume that $\Psi_{T_i}$ is square integrable for all $M-1<i\leq N$.

The price of the option can be calculated by applying the above recursion up to $T_M$, before taking the expectation under $\mathds{Q}$.
\begin{equation}
 V_0 := \ExO{ \frac{V_{T_M}}{B_{T_M}} }.
\end{equation}%

\section{Pricing techniques}
\label{sec:pricing}

This section presents the different pricing techniques we will investigate in our numerical experiments later on in Section~\ref{sec:numerics}. In the case of floating-strike Asian options we will use also two methods proposed in~\textcite{Goudenege2022}. The former is based on Gaussian Process Regression and Gauss-Hermite quadrature, and it is named GPR-GHQ. The latter is based on the construction of a binomial Markov chain. It is beyond the scope of the present article to adapt their methodologies to other payoffs. We refer to Appendix~\ref{app:Zanette} for a description of the main features of the two methodologies.

In the dynamic problem of Equation~\eqref{eq:asian_dynprog} a choice is made between an intrinsic value, $\Psi_{T_i}$, and a continuation value, the latter being defined by a conditional expectation as
\begin{equation}
\label{eq:continuation}
  C_{T_i} := B_{T_i}\ExF{T_i}{\frac{V_{T_{i+1}}}{B_{T_{i+1}}} }
\end{equation}%
with $M-1<i<N$; the continuation value is, in general, not trivial to calculate. In addition, we notice that it is not only a function of the last stock price, say $S_{T_i}$, but a function that potentially depends on the entire history $S_{T_{M}}, \ldots, S_{T_i}$. In what follows, it may happen that we will denote explicitly the dependence of the continuation value on the information set if necessary.

Now, we proceed by describing how we approximate the continuation values $C_{T_i}$ given by Equation~\eqref{eq:continuation}.

\subsection{Least-square Monte Carlo}
\label{sec:lsmc}

A standard way to approximate the continuation values given by Equation~\eqref{eq:continuation} is the LSMC described by~\textcite{Tilley1993,Barraquand1995,Longstaff2001}. The LSMC method re-writes the dynamic problem given by Equation~\eqref{eq:asian_dynprog} as
\begin{equation}
V_{T_i} = \ind{\Psi_{T_i}>C_{T_i}} \Psi_{T_i} + \ind{\Psi_{T_i}<C_{T_i}} C_{T_i} .
\end{equation}%
First, we approximate the continuation values occurring within the indicator functions as
\begin{equation}
\label{eq:continuation_approx}
 C_{T_i} \approx C^\theta_{T_i} := \sum_{b=1}^B \theta_b \varphi_b(S_{T_1}, \ldots, S_{T_i})
\end{equation}
where $\{\varphi_1,\ldots,\varphi_B\}$ is a set of $B$ basis functions and $\theta \in \mathds{R}^{B}$ are the trainable weights, which can be estimated by standard linear regression. Then, if we expand the recursion, we notice that the extant conditional expectation at time $T_i$ is evaluated within expectations conditioned at previous times, up to time $T_M$. Thus, the dynamic problem can be formulated in an equivalent way, thanks to the tower rule, as
\begin{equation}
\label{eq:asian_ls}
 V_0 \approx \ExO{\frac{V^\theta_{T_M}}{B_{T_M}}} , \quad
 V^\theta_{T_i} := \ind{\Psi_{T_i}>C^\theta_{T_i}} \Psi_{T_i} + \ind{\Psi_{T_i}<C^\theta_{T_i}} \frac{B_{T_i}}{B_{T_{i+1}}} \,V^\theta_{i+1}, \quad
 V^\theta_{T_N} := \Psi_{T_N}.
\end{equation}%
We warn the reader that $V^\theta_{T_i}$ does not represent the option value at time $T_i$ since it depends on future realization of the price process. The option value is its expectation conditioned at time $T_i$. 

Different types of basis functions have been proposed in the literature so far. In the original paper of~\textcite{Longstaff2001}, the authors discuss Laguerre, Hermite, Legendre, Chebyshev, Gegenbauer, and Jacobi polynomials. A Laguerre polynomial approximation is used also in~\textcite{bernhart2011finite} to price a continuously monitored moving average options. In his Ph.D. thesis, \textcite{grau2008applications} employs a sparse polynomial basis for the regression in order to attenuate the numerical inefficiency of the LSMC method. However, the latter does not easily scale to high-dimensional problems. \textcite{dirnstorfer2013high} use sparse grid basis functions based on polynomials or piece-wise linear functions. \textcite{lelong2019pricing} replaces the standard least-square regression with a Wiener chaos expansion. Finally, we mention the Ph.D. thesis of~\textcite{bilger2003valuation}, where polynomials of the underlying index and the average are employed. We will see in Section~\ref{sec:numerics} that the choice of the basis functions will have a non-negligible impact on the pricing performances.

\subsection{Randomized neural networks}
\label{sec:rnn}

We employ two types of approximations of the continuation value based on randomized neural networks. The use of randomized neural networks for computing the price of American options has been firstly introduced in~\textcite{herrera2021optimal}, with impressive results in terms of speed and accuracy.

In fact, this method allows to build a random basis that is rich enough for a good fit at a cheap computational cost, and just train the last layer for the network, which amounts to just solve a linear regression (at most with regularization). In this way, randomized neural networks can be viewed as an alternative to the polynomial bases, usually adopted when implementing the LSMC method, which can better represent non-linearity and path-dependency of the continuation value. The first type of randomized neural network approximates the continuation function via a randomized feed-forward neural network; we call this type of networks R-FFNN. On the other hand, the second type employs a randomized recurrent neural network for the approximation of the continuation value; we call this type of network R-RNN. 

In the upcoming subsections, we follow the presentation by~\textcite{herrera2021optimal} for the description of the two methodologies; in particular we present them with a fully-connected shallow neural network.

\subsubsection{R-FFNN}
\label{subsec:r-ffnn}

We start by introducing the activation function $\sigma\,:\,\mathds{R}\rightarrow\mathds{R}$ acting on each component of the network, for instance a rectified linear unit (ReLU) or an hyperbolic tangent (tanh) activation function. Let $\widetilde{M}, h \in \mathds{N}_{>0}$, we define $\boldsymbol{\sigma} \;:\; \mathds{R}^{h-1}\rightarrow\mathds{R}^{h-1}$ as the activation function acting component-wise, in the sense that $\boldsymbol{\sigma}(x)=(\sigma(x_1),\ldots,\sigma(x_{h-1}))^{\top}$, where $x \in \mathds{R}^{h-1}$. The natural number $\widetilde{M}$ depends on the choice of the information set used for the construction of the random basis. For instance, in case of Asian options (see Subsection~\ref{sec:amerasian}), one possibility is to use the values of the stock price over the time window of length $M$; in this case $\widetilde{M}=M$.

Then, let $d \in \mathds{N}_{\geq 1}$ be the dimension of of the state process at each observation date, we introduce the parameters of the hidden layer $\vartheta:=(A, b) \in \mathds{R}^{(h-1)\times(d\widetilde{M})} \times \mathds{R}^{h-1}$, namely the weight matrix $A$ and the bias vector $b$, whose elements are randomly and identically sampled and not optimized in any training procedure (we will specify their distribution at the beginning of Section~\ref{sec:numerics}). Furthermore, we introduce the function $\phi:\mathds{R}^{d \widetilde{M}}\rightarrow\mathds{R}^{h},\,x\mapsto\phi(x) = (\boldsymbol{\sigma}(A x + b)^\top,1)^\top$ and the parameters $\theta_{i}:=((A_{i})^\top,b_{i}) \in \mathds{R}^{h-1}\times\mathds{R}$ to be optimized. We notice that the element $1$ in $\phi$ which is the constant element that is added for standard linear regressions.

Finally, we are able to formulate the approximation of the the continuation value. We can write for each $i$
\begin{equation*}
  C_{i}^{\theta}(x):=\theta_{i}^\top\phi(x) = A_{i}^\top\boldsymbol{\sigma}(A x + b) + b_{i}.
\end{equation*}%
where $\theta_{i}$ can be estimated via OLS because the approximation function $C_{i}^{\theta}$ is linear in the parameters $\theta_i$; see Theorems~5 and~6 of~\textcite{herrera2021optimal}.

We can exemplify the above construction by analyzing a specific case. For instance, for the sake of exposition only, we suppose that $M=1$, and that we have access to $m$ realizations of the stock price paths, where the $\ell$-th realization is denoted by $S_0, S_{T_1}^{(\ell)}, S_{T_2}^{(\ell)}, \ldots, S_{T_N}^{(\ell)}$ with the fixed initial value $S_0$. Then, the parameters $\theta_i$ of the last layer are found by minimizing the squared error of the difference between conditional expectation of the discounted future price of the financial derivative, say $\frac{B_{T_i}}{B_{T_{i+1}}} V_{T_{i+1}}^{\theta, (\ell)}$ and the approximation of the continuation value evaluated at $S_{T_i}^{(\ell)}$. The minimizer has the following closed form expression:
\begin{equation*}
    \theta_i = \frac{B_{T_i}}{B_{T_{i+1}}} \left(\sum_{\ell=1}^{N_{MC}} \phi\left(S_{T_i}^{(\ell)}\right)\phi^{T}\left(S_{T_i}^{(\ell)}\right)\right)^{-1} \cdot \left(\sum_{\ell=1}^{N_{MC}} \phi\left(S_{T_i}^{(\ell)}\right) V_{T_{i+1}}^{\theta, (\ell)}\right)
\end{equation*}
The fact that the training of the previous model (more generally, of all the models described in the present paper) can be performed in a single analytical step, that is implemented efficiently in most linear algebra libraries, is the major advantage of these methods. The computational complexity of the least square method is mostly influenced by the $h \times h$ matrix inversion in the previous equation.

One limitation of the R-FFNN is that all the points in the information set have the same relevance, in the sense that the fact that the input may be a stream of data is not captured by the R-FFNN. The R-RNN in the next paragraph takes into account this information. 

\subsubsection{R-RNN}
\label{subsec:r-rnn}

In the case of a R-RNN the parameters of the hidden layer, which are randomly and identically sampled and not optimized, are given by $\vartheta:=(A_x, A_\xi, b) \in \mathds{R}^{(h-1)\times (d\widetilde{M})} \times \mathds{R}^{(h-1) \times (h-1)} \times \mathds{R}^{(h-1)}$. We notice that, in this case, the tuning parameters are more important, as they determine the interplay between past and new information.

We start by defining $\phi: \mathds{R}^{d \widetilde{M}}\times\mathds{R}^{h}\rightarrow\mathds{R}^{h+1}$, $(x,\xi)\mapsto\phi(x,\xi)=$ $(\boldsymbol{\sigma}(A_x x + A_\xi \xi + b)^\top, 1)^\top$, and $\theta_{i}:=((A_{i})^\top, b_{i}) \in \mathds{R}^{h-1}\times\mathds{R}$ the parameters to be optimized. For each $i$ the continuation value is approximated by
\begin{equation*}
  \begin{cases}
    \xi_{i} = \boldsymbol{\sigma}(A_x x_{i} + A_{i} \xi_{i-1} + b),\\
    C_{i}^{\theta}(x) =\theta_{i}^\top\phi(x) = A_{i}^\top\boldsymbol{\sigma}(A x + b) + b_{i} = \theta_{i}^\top\phi(x_{i}, \xi_{i-1}),
  \end{cases}
\end{equation*}%
where $\xi_{-1}=0$. As for the R-FFNN the parameters $\theta_{i}$ are estimated via OLS, and so the computational complexity of the algorithm is mostly influenced by the inversion of a $(h-1) \times (h-1)$ matrix. 

There are a number of hyper-parameters for both R-FFNN and R-RNN that can be tuned, namely the size of the neural network hidden state, the number of layers in the neural network, the activation function, and the distribution of the parameters of the hidden layer. For the R-RNN, we consider one layer in time for every fixing date. From a practical point of view, the main difference between R-FFNN and R-RNN is that for the former we construct a new neural network in which the number of layers coincides with the number of fixing dates, whereas for the R-RNN we construct a unique (random) neural network in which we train each time only the readout map. In particular, the dimension of the R-RNN decreases going backward in time.

\subsection{Signature methods}
\label{sec:signature}

We propose a novel alternative algorithm to R-FFNN and R-RNN based on signatures. We design the algorithm to preserve the linear regression on which is based the LSMC method, so that we can consider also the following algorithm as an alternative definition of the basis function. More precisely, we analyze two versions of the algorithm: the former one based on the truncated signature of a suitable transformation of the time series of observed asset prices, the second one on the randomized signature. Non-randomized signatures are, roughly speaking, a tool that allows to extract features and summarize a stream of information over a time interval. However, they may suffer from similar disadvantages as polynomials, such as the ``curse of dimensionality'', which can be overcome by randomized signatures.

\subsubsection{Truncated signature}

The signature process was first studied by~\textcite{chen1957integration,chen1977iterated}, and plays a prominent role in rough-path theory introduced in the~\textcite{lyons1998differential}. In an effort to keep the paper as self-contained as possible we provide for convenience in Appendix~\ref{app:signature}, Section~\ref{app::standardsignature}, a self-supporting although minimalist account of key concepts and results on signatures. Here, instead, we give some results on signatures, related to the present work, within a univariate time series framework; they can be easily generalized to a multi-dimensional time series. The material is based on~\textcite{levin2013learning}, Section~4, and~\textcite{gyurko2013extracting}.

For any $0 \leq i_1 < i_2 \leq N$, and $i_2, i_1 \in \mathds{N}$, in general the signature of $\{(T_j, S_{T_j})\}_{j=i_1}^{i_2}$ is computed via the following two steps:
\begin{itemize}
  \item[(i)] embed a time series $\{(T_j, S_{T_j})\}_{j=i_1}^{i_2}$ into a continuous path $R$;
  \item[(ii)] compute the signature of this transformed continuous path $R$.
\end{itemize}
We add time as one of the variables in the time series $\{(T_j, S_{T_j})\}_{j=0}^{N}$ to preserve the temporal structure; this also ensures that signatures are unique. This procedure is known as time augmentation in~\textcite{hambly2010uniqueness} (see Section~\ref{app:numerical_aspects} for more details).

There are several choices for embedding a time series into a continuous path. For instance, one can rely directly on a simple piece-wise linear interpolation technique. However, the signature is an infinite sequence, whence in practice some finite collection of terms must be selected. 
Because the magnitude of the terms exhibit factorial decay, it is usual, as in~\textcite{lyons2014rough}, to simply choose the first $n$ terms of the sequence and consider the so-called \textit{truncated signature} of depth $n$. Yet, if the function to be learned depends non-trivially on the higher degree terms, then crucial information has nonetheless been lost. A remedy might be using embedding techniques which, roughly speaking, apply a point-wise augmentation to the original stream of data before taking the signature. Then, the first $n$ terms of the signature may better encode the necessary information.

On a first glance, the embedding transformation could be seen as not strictly necessary, since it duplicates the (spatial) dimension of the time-series. In order to answer this question, in our numerical Section~\ref{sec:numerics}, we compare the performance of the signature method with and without embedding transformations.

In the present paper, we employ, sequentially, two embedding techniques  that we define in the following: (a) the Hoff lead-lag transformation (\textcite{hoff2006brownian} and~\textcite{flint2016discretely}), and (b) the time-joined path.

\begin{definition}[Hoff lead-lag transformation]
Let $\{(T_j, S_{T_j})\}_{j=0}^{N}$ be a univariate time series. Let $Z\,:\,[0,2N]\rightarrow\mathds{R}^{+}\times\mathds{R}$ be a 2-dimensional lead-lag transformation of $\{(T_j, S_{T_j})\}_{j=0}^{N}$, which is defined as follows:
\begin{equation}
\label{eq:leadlag}
  Z(t):=
  \begin{cases}
    S_{T_j}e_1 + S_{T_{j+1}}e_2\quad\quad\quad\quad\quad\quad\quad\quad\quad\quad\quad\quad\quad\quad\,\,\,\,\,\,\text{if}\,\,t\in[2 j, 2 j+1)\\
    S_{T_{j}}e_1 + [S_{T_{j+1}}+2(t-(2j+1))(S_{T_{j+2}}-S_{T_{j+1}})]\,\,\,\,\,\,\,\,\text{if}\,\,\,\,t\in[2 j+1, 2 j+\frac{3}{2})\\
    [S_{T_{j}}+2\left(t-\left(2i+\frac{3}{2}\right)\right)\left(S_{T_{j+1}}-S_{T_{j}}\right)]e_1+S_{T_{j+2}}e_2\,\,\,\,\text{if}\,\,t\in[2 j+\frac{3}{2}, 2 j+2)\\
  \end{cases}
\end{equation}%
for $t\in [0, 2N]$, and where $\{e_i\}_{i=1,2}$ is an orthonormal basis of $\mathds{R}^2$.
\end{definition}

\begin{definition}[Time-joined transformation]
Let $\{(T_j, S_{T_j})\}_{j=0}^{N}$ be a univariate time series. Let $R\,:\,[0,2N+1]\rightarrow\mathds{R}^{+}\times\mathds{R}$ be a 2-dimensional time-joining path of $\{(T_j, S_{T_j})\}_{j=0}^{N}$, which is defined as follows:
\begin{equation*}
  R(t):=
  \begin{cases}
    T_{0} e_1 + S_{T_0}te_2,\quad\quad\quad\quad\quad\quad\quad\,\,\quad\quad\quad\quad\quad\,\,\,\,\,\,\,\,\,\text{if}\,\,t\in[0,1);\\
    [T_j+(T_{j+1}-T_{j})(t-2 j -1)]e_1 + S_{T_j} e_2\quad\quad\,\,\,\,\,\,\,\text{if}\,\,t\in[2j+1,2j+2);\\
    T_{j+1}e_1+[S_{T_j}+(S_{T_{j+1}}-S_{T_j})(t-2j-2)]e_2,\quad\,\,\text{if}\,\,t\in[2j+2, 2j+3);\\
  \end{cases}
\end{equation*}%
where $0\leq j \leq N-1$, $t\in[0,2N+1]$ and $\{e_i\}_{i=1,2}$ is an orthonormal basis of $\mathds{R}^2$.
\end{definition}

In particular, the signature of $\{(T_j, S_{T_j})\}_{j=0}^{N}$ is defined to be the signature of its time-joined transformation $R(t)$. There also exists a more intuitive and straightforward definition of a lead-lag transformation. Yet, we decide to consider the so-called Hoff lead-lag transformation because its good performance has been already noticed in the literature across different data sets and algorithms; see, for instance, \textcite{fermanian2021embedding}.

An advantage of the lead-lag transformation is that one can read the volatility of the path directly from the second term of the signature; see Remark~4.1 of~\textcite{levin2013learning}. On the other hand, the time-joined transformation, is such that the resulting continuous path exhibits a causal dependence on the data, being an hybrid between a linear and a rectilinear interpolation; see~\textcite{moor2020path}.

In the present work, we use truncated signatures as non-linear basis on which we can estimate linear regressions to have theoretically sounded results (see Theorem~\ref{thm: sig-universal-approx}). In particular, for each realization of the stock price over the time window $T_0, \ldots, T_N$, we first apply, sequentially, the lead-lag transformation and the time-joined transformation to $\{(T_{j}, S_{T_j})\}_{j=T_{i_1}}^{T_{i_2}}$, then we apply the truncated signature of order $n$ to the latter. We notice that by performing such transformations, every transformed path is enriched with other dimensions: the lead-lag transform will double the dimension of the original time series, while time-addition will increase it by one. In particular, the previous two transformations, applied in sequence, transform a univariate time series in a $d=3$ dimensional time series. Only afterwards, one computes the signature. The output of the truncated signature is a vector of dimension $s_d(n) = \frac{d^{n+1}-1}{d-1}$; see~Appendix~\ref{app:signature}. We use this vector, by eventually enlarging it with, for instance, the average of the stock price over the considered time window, as linear regression basis for the approximation of the continuation value; a regularization can be applied to the linear regression.

The approach explained in this section for American-style Asian options has been used in~\textcite{Feng2021sig,Bayraktar2022}, but, in this case, the signature transformation is applied to the path and then fed to a feed-forward neural network, which should not be strictly necessary.

The drawback of using the truncated signature of order $n$ is that for a sample consisting of $N$ points in $\mathds{R}^{d}$ it has an $\mathcal{O}(N d^{n})$ computational complexity, which becomes intractable for high-dimensional systems and/or for large values of $n$ aimed at encoding a richer information of the function to be learned; the complexity is, however, linear in the number of sampled points. In order to cope with the just-mentioned issue, instead of calculating the signature of a time series, one can extract a new quantity, called randomized signature, which is easier to compute and inherits the expressiveness of the signature. This is the content of the next subsection.

\subsubsection{Randomized signature}
\label{sec:rand_signature}

Randomized signature has been proposed as a flexible and easily implementable alternative to the well-established path signature. Despite the name, the concept of randomized signature is strictly connected to that of reservoir computing, an area of machine learning where random, possibly recurrent neural networks are used to efficiently construct regression basis on path space. In what follow, we give a brief overview of reservoir based on the randomized path signature; we first explain the notion of reservoir system, then we present signatures as reservoirs.

Randomized signatures emerged by considering signatures as reservoir systems, that is algorithms that try to approximate input/output systems without the need of a full calibration of the system itself; see, for more details, \textcite{Cuchiero22_randSig} and the references therein. A universal approximation property is granted also in this case. This intuition was first developed in~\textcite{Cuchiero22_randSig} and then in~\textcite{Cuchiero_Gonon_Grigoryeva_Ortega_Lahuerta_Teichmann_2021}, obtaining what the authors called ``randomized signatures'', where the approximation property relies on the Johnson-Lindenstrauss Lemma (see~\textcite{Johnson1984ExtensionsOL}), consisting in a random projection of the system on a space with smaller dimension that is able to preserve its geometric structure: distortion will be at most $1+\epsilon_r$ for arbitrary $\epsilon_r$, and the embedded space will also have dimension that depends on the same $\epsilon_r$.

One notable drawback of considering signatures as reservoirs is that the computation of the iterated integrals needed to form the basis of the path is rather lengthy and computationally intense. This problem becomes even more relevant in high dimensions or for very long time series. It is indeed in such cases that randomized signatures, leveraging a random construction, can make a difference. The idea behind randomized signature is compressing information of the signature through the Johnson-Lindenstrauss (JL) Lemma and consider a dynamical system for this projected signature-transformed stream.

\begin{lemma}[\textcite{Johnson1984ExtensionsOL}]\label{thm: JL-lemma}
  For every $0 < \epsilon_r < 1$, an $N$ point set $Q$ in some arbitrary (scalar product) space $\mathds{H}\subseteq \mathds{R}^d$ can be embedded into a metric space $\mathds{R}^{k^\star}$, where $k^\star = \frac{24 \log(N)}{3\epsilon^2_r-2\epsilon^3_r} = \mathcal{O}(\frac{\log N}{\epsilon^2_r})$ in an almost isometric manner, i.e.~there is a linear map $g: \mathds{H} \to \mathds{R}^{k^\star}$ such that $(1-\epsilon_r)\| p_1-p_2\|^2 \leq \| g(p_1)-g(p_2)\|^2 \leq (1+\epsilon_r) \| p_1-p_2\|^2$ for any couple of points $p_1$ and $p_2$ in $Q$. The map $g$ can be chosen to be random.
\end{lemma}

\noindent We have the following definition (see, e.g.,~\textcite{Effectiveness_RandSig}):

\begin{definition}[Randomized signature]
  Given $k^{\star} \in \mathds{N}$ and random matrices $A_1,\ldots,A_d \in \mathds{R}^{k^{\star} \times k^{\star}}$, random shifts $b_1, \ldots, b_d$ in $\mathds{R}^{k^{\star}}$, and any fixed activation function $\sigma$, the randomized signature of a continuous and piece-wise smooth path $X : [0,T] \rightarrow \mathds{R}^{d+1}$ is the solution of the following differential equation
  \begin{equation}
  \label{eq:ran_sig}
    d \mathrm{RS}(X)_t = \sum_{i=0}^{d} \boldsymbol{\sigma}(A_i \mathrm{RS}(X)_t+ b_i) \,dX^i_{t},\quad \mathrm{RS}(X)_0 \in \mathds{R}^{k^\star}.
  \end{equation}
  Notice that we typically take $X_t^{0}=t$.
\end{definition}

In particular, $A_i$ and $b_i$ can be sampled from a normal distribution and are fixed once for all before applying the Euler scheme to Equation~\eqref{eq:ran_sig}. More precisely, once that a dimension $k^{\star} \in \mathds{N}$ is fixed, every entry of the random matrices can be chosen independently from a normal distribution $\mathcal{N}\left(0,\varsigma^2\right)$. We also notice that the function $\boldsymbol{\sigma}$, which might remind the reader to the activation function of neural networks, does not need to be non-linear and can be chosen to be the identity. Most importantly, the activation function needs to be injective to obtain a non-singular $\boldsymbol{\sigma}(A_i \mathrm{RS}(X)_t+ b_i)$ for any $i=1, \dots, d$. For a discussion on the possible assumptions on the activation functions, we refer the reader to \textcite{Akyildirim2022anomaly} and \textcite{biagini2024universal}. This is not the only viable definition of randomized signatures based on the JL lemma. Another possible definition is present in the work by Cuchiero and M\"{o}ller (\textcite{Cuchiero22_randSig}, Definition 3.8) where the distinction between the application of randomized linear functionals from the JL lemma to truncated signature coefficients is seen as an alternative to considering randomized signatures as solution to Equation~\eqref{eq:ran_sig}. The former are thus called \emph{JL-signatures}, while the latter \emph{randomized signatures}. The definition we give here is more in line with the reservoir systems' literature outlined in previous paragraphs.

Examples of applications of randomized signatures can be found, for example, in~\textcite{Akyildirim2022anomaly}, where they are used to identify pump and dump attempts in the crypto market in an unsupervised learning settings, and in~\textcite{Akyildirim2023portfolio}, where randomized signatures are utilized as a non-parametric and non-linear drift estimator to find an optimal allocation a long-only portfolio.

The computational complexity for calculating $\mathrm{RS}$ in Equation~\eqref{eq:ran_sig} is $\mathcal{O}(k^{\star^2} d)$ and at any time its dimension is $\mathcal{O}(k^{\star})$ (when solving Equation~\eqref{eq:ran_sig} numerically, the dimension is also proportional to the number of time steps used for the Euler scheme). 
In practice, $k^\star \in \mathds{N}$ is normally chosen as a parameter of the method and its value must be commensurate with the expressiveness we require from the algorithm. 
In Section~\ref{sec:numerics} we show experimentally that, in order to match the pricing capabilities of the truncated signature of order $n$, the number $k^{\star}$ of required randomized signatures is fairly small, thus leading to a dimension that is way smaller than the dimension of the standard signature $s_d(n)$ introduced in the previous section. 
This confirms that working with randomized signatures is often less computationally demanding.

\subsection{Sensitivity computation}
\label{sec:sensitivity}

The pricing algorithm we are discussing for early-termination products require the computation of a regression at each time step of the simulation algorithm. This calculation introduces additional noise in the results, so that the numerical evaluation of sensitivities to market parameters could be unstable. In particular, this can be the case for second-order sensitivities.

We consider the case of calculating first and second order sensitivities with respect to the initial spot price (namely we are interested in the so-called Delta and Gamma), and we start to analyze the simpler case of American put options, since in this case we have the possibility to check our methods against a standard (and reliable) implementation based on a binomial tree. We start by discussing the problems of finite-difference schemes, then we introduce two alternative methods: (i) the regression method proposed in~\textcite{Letourneau2019}, and later used in~\textcite{herrera2021optimal}, and (ii) the proposal of~\textcite{Maran2020} based on Chebyshev interpolation. We extend the latter on to our specific payoff case.

\subsubsection{Finite-difference method}
\label{sec:fd}

The simplest, and more direct, approach for sensitivities is the discretization of derivative operators in term of finite differences. If we consider a positive number $\epsilon$, we can approximate Delta and Gamma as given by
\begin{equation}
\frac{\partial V}{\partial S} \approx \frac{V(S(1+\epsilon))-V(S(1-\epsilon))}{2\epsilon}
\;,\quad
\frac{\partial^2 V}{\partial S^2} \approx \frac{V(S(1+\epsilon))-V(S)+V(S(1-\epsilon))}{\epsilon^2}
\end{equation}%
for small values of $\epsilon$, where $V(s)$ is the option price calculated with spot price equal to $s$.

\subsubsection{Regression method}
\label{sec:regression}

We briefly describe the regression method applied to our case, but we address the reader to the original paper by~\textcite{Letourneau2019} for more details and for a more general exposition. We assume that for a given positive $\epsilon$ the initial spot price $S_\epsilon$ of the Monte Carlo simulation is distributed according to a normal random variable ${\cal N}(S,\epsilon^2)$, where $S$ is the market spot price, so that the derivative price can be written as
\begin{equation}
 V = \ExO{V(S_\epsilon)} \approx \frac{1}{N_s} \sum_{k=1}^{N_s} {\hat V}(S^{(k)}_\epsilon)
 \;,\quad
 S_\epsilon \sim {\cal N}(S,\epsilon^2) ,
\end{equation}
where $N_s$ is the number of simulation paths and $S^{(k)}_\epsilon$ is the $k$-th extraction of the random variable $S_\epsilon$. Such extraction is used as initial spot price for the Monte Carlo approximation ${\hat V}(S^{(k)}_\epsilon)$ of the derivative price made with $N_s$ simulation paths.

The previous decomposition allows us not only to calculate the derivative price, but also its sensitivities. Indeed, we can consider the Taylor's approximation
\begin{equation}
 {\hat V}(x) \approx \sum_{b=0}^B \beta^{(\epsilon,B)}_b (x-S)^b ,
\end{equation}
where the coefficients $\beta$ can be calculated by means of an OLS algorithm on the  data set $\{(S^{(k)}_\epsilon,{\hat V}(S^{(k)}_\epsilon))\}_{k=1}^{N_s}$. Then, the sensitivities can be calculated as given by
\begin{equation}
\frac{\partial V}{\partial S} \approx \beta^{(\epsilon,B)}_1
\;,\quad
\frac{\partial^2 V}{\partial S^2} \approx 2\beta^{(\epsilon,B)}_2
\end{equation}

\subsubsection{Chebyshev method}
\label{sec:chebyshev}

The method of~\textcite{Maran2020}, hereafter termed Chebyshev method, consists in approximating the option price, view as a function of the spot price, by means of a Chebyshev interpolator. Then, sensitivities are calculated by taking the derivative of the interpolator with respect to the spot price. The technique relies on the fact that for analytical functions the rate of convergence of the derivatives of the interpolator to the original derivatives is exponential. The strength of the method relies on the fact that, thanks to the previous property, we can increase the size of the interpolation interval to stabilize the sensitivity computation without introducing additional biases. More details on using Chebyshev interpolation techniques in option pricing can be found in~\textcite{Gass2016}.

Here, we present our procedure in details. For the sake of clarity we skip the technical part of dealing with Gamma discontinuities, which we will analyze in Appendix~\ref{app:sensitivity}. For a given positive $\epsilon$ we select a number $N_c$ of spot prices $\{S^0_\epsilon,\ldots,S^{N_c-1}_\epsilon\}$ where we evaluate the corresponding Monte Carlo approximations $\{{\hat V}(S^0_\epsilon),\ldots,{\hat V}(S^{N_c-1}_\epsilon)\}$ of the derivative price. Such spot prices can be defined as
\begin{equation}
S^\ell_\epsilon := S \left(1 + \epsilon \cos\left(\frac{\ell\pi}{N_c-1}\right)\right) ,
\label{eq:chebpoint}
\end{equation}%
where $S$ is the market spot price. The set of points $\{(S^\ell_\epsilon,{\hat V}(S^\ell_\epsilon))\}_{\ell=0}^{N_c-1}$  will be used to train a polynomial interpolator given by
\begin{equation}
{\hat V}_\epsilon(S) := \sum_{\ell=0}^{N_c-1} {\hat V}(S^\ell_\epsilon) \,\prod_{\jmath\ne \ell}^{N_c-1} \frac{S - S^\jmath_\epsilon}{S^\ell_\epsilon - S^\jmath_\epsilon} ,
\end{equation}
The interpolator allows us not only to calculate the derivative price, but also its sensitivities by explicit calculation of the derivatives with respect to the spot price.

\subsubsection{Algorithm comparison for American put Gamma}
\label{sec:sensitivity-americanput}

We wish to assess the performance of these methods on early-termination products before proceeding with their use in the in the numerical Section~\ref{sec:numerics}. We consider as a typical playground the calculation of the second derivative with respect to the spot price (Gamma) for American put options. All the calculations are made according to the Black and Scholes model, later described in Section~\ref{sec:dynamics}.

Each method to compute the sensitivities may require a different number of Monte Carlo simulations; for instance, as shown above, finite differences require three different simulations to compute Gamma. Thus, in order to compare the resulting sensitivities, keeping fixed the computational time, we consider the same number of simulated paths for each approach, irrespectively of how they are distributed among the required simulations.

We list in Table~\ref{tab:gamma_mc} the results obtained for Gamma computation for the three methods for three different choices the spot price, while keeping $K=100$ in all calculations. We compare these results with the Gamma calculations made by means of a binomial tree. We can see, as the number of simulation paths increases, that the regression method quickly deteriorates by showing a significative bias. The other two methods perform with greater accuracy. In particular, the Chebyshev method shows a smaller bias.

These results suggest us to discard the regression method, and focus only on finite-difference and Chebyshev methods, in the numerical investigations we are going to illustrate in the following sections.

\begin{table}
 \begin{center}
  \pgfplotstabletypeset[
    string replace={0}{},
    fixed, fixed zerofill, precision=4, set thousands separator={},
    every head row/.style={after row=\midrule,before row=
       \toprule
       \multicolumn{10}{l}{\bf American put option: Gamma} \\
       \midrule
       & \multicolumn{3}{c|}{$\bm{\Gamma=0.0235,\, S_0=85}$}
       & \multicolumn{3}{c|}{$\bm{\Gamma=0.0308,\, S_0=100}$}
       & \multicolumn{3}{c}{$\bm{\Gamma=0.0131,\, S_0=115}$} \smallskip\\
       },
    every last row/.style={after row=\bottomrule},
    columns={npath,e085,r085,c085,e100,r100,c100,e115,r115,c115},
    columns/npath/.style={sci, precision=1, column name={\bf MC Path}},
    columns/e085/.style={column name={\bf F.Diff.}, column type/.add={|}{}},
    columns/r085/.style={column name={\bf Regr.}},
    columns/c085/.style={column name={\bf Cheb.}},
    columns/e100/.style={column name={\bf F.Diff.}, column type/.add={|}{}},
    columns/r100/.style={column name={\bf Regr.}},
    columns/c100/.style={column name={\bf Cheb.}},
    columns/e115/.style={column name={\bf F.Diff.}, column type/.add={|}{}},
    columns/r115/.style={column name={\bf Regr.}},
    columns/c115/.style={column name={\bf Cheb.}},
  ]{\sensmc}
 \end{center}
 \caption[Gamma for an American put option]{Gamma for an American put option with maturity $T=0.2$, strike $K=100$ and time steps $N=50$. The first column is the total number of simulation paths used by each algorithm. Then, there are three groups, each of three columns, for different spot prices $S_0$. In each group, the first column is the result of the finite-difference method ($\epsilon=1/16$), the second one of the regression method ($\epsilon=1/8$, $B=9$) and the third one of the Chebyshev method ($\epsilon=1/8$). In the second line $\Gamma$ is the result of the binomial tree.}
 \label{tab:gamma_mc}
\end{table}

\section{Numerical techniques}
\label{sec:numerics}

This section describe the ingredients common to all the payoffs and algorithms. In particular, we discuss the underlying asset dynamics, the input features of the regression basis functions, and the configuration of the random networks and signature methods.

\subsection{Price dynamics}
\label{sec:dynamics}

Since our contribution is focused on analyzing the performance of selecting different basis functions in the LSMC algorithm, and how this selection depends on the specific payoff we consider, we think that assuming a simple setting for asset price dynamics is preferable. In particular, we evaluate the different algorithms by assuming a Black-Scholes dynamics for underlying assets. Admittedly, we could have used more complicated dynamics, such as stochastic volatility models or rough volatility models. Moreover, we only consider synthetic data, and not real market data. However, these extensions would not add to the present work's conceptual advancements

The Black-Scholes model (see~\textcite{black1973pricing}) is characterized by the following stochastic differential equation:
\begin{equation}
\label{eq:blackandscholes}
  dS_t = (r-q) S_t\,dt + \sigma S_t\,dW_t,
\end{equation}%
where $t \in [0,T]$, $S_0=s_0$, and $(W_t)_{t \geq 0}$ is a standard Brownian motion. Besides, $r$ is the risk-free rate, $q$ is the dividend rate, and $\sigma$ is the volatility, and we assume that all these three quantities are deterministic constants.

In our numerical investigation we use the following values: $S_0=100$, $r=0.05$, $q=0$, $\sigma=0.3$. Such values are used also in~\textcite{bernhart2011finite} and (later on) in~\textcite{Goudenege2022}.

\subsection{Risk factors and features}
\label{sec:riskfactor}

In the following, we compare four different types of basis functions for the LSMC algorithm: (i) polynomials, (ii) R-FFNN, (iii) R-RNN, (iv) signatures (either randomized or not).

As shown in Equation~\eqref{eq:continuation_approx} the notation $\varphi_b(S_{T_1}, \ldots, S_{T_i})$, with $1\le b\le B$, is used to state that basis functions can depend on any value of the price process up to regression time $T_i<T_N$. However, depending on the payoff under consideration and on the choice of the basis functions, we can refine this statement.

In our case of Asian and look-back payoffs, we should consider at each regression time $T_i$ the observations of the underlying asset on all the previous $M-1$ days, where $M$ is the number of days of the moving window. Thus, at each regression time $T_i$ we should consider basis functions of the form $\varphi_b(S_{T_{i-M+1}}, \ldots, S_{T_i})$. This can be simply accommodated in the polynomial case by considering monomials within a given degree in the arguments. Yet, these methods have in principle the limitation that they assign the same relevance to all the points in the information set, in the sense that they do not capture, for instance, the fact that they are a stream of data, namely there is temporal dependence among the observation of the price process. On the other hand, R-RNN and signature methods take into account this fact.

In the case of polynomial and R-FFNN basis functions when the length $M$ of the moving window increases the performance of the algorithm deteriorates, since we need either to increase the number of basis functions or of features of the network. As a possible solution to this problem we can reduce the number of observations needed to characterize the input and we can check empirically the impact on options prices. 

With this considerations in mind we proceed as following for the polynomial and R-FFNN cases. As denoted above, let $M$ be the number of observed underlying values included in the window-average (specific to the option derivative) and $T_i$ the evaluation time of the regression. We build our basis functions by starting from four different choices of sets of risk factors, which we denote $\risk{\rho}$ with $\rho\in\{1,2,3,4\}$. Each risk factor set contains $F^\rho_i$ random variables according to the following rules.
\begin{itemize}
\item {[$\risk{1}$]} \label{itm:B1} We consider only one risk factor, namely the value of the stock price $S_{T_i}$ at time $T_i$. This choice is a strong approximation, since it discards all path-dependency effect. On the other way, is the maximum speed up we can obtain.
\item {[$\risk{2}$]} \label{itm:B2} We consider two risk factors (only for $M>1$, in case $M=1$ we revert to $\risk{1}$): the value of the stock price $S_{T_i}$ at time $T_i$ and the arithmetic moving average computed over the interval $[T_{i-M+1}, T_{i}]$. Also this choice is a strong approximation, but it tries to synthesize all the path-dependency effects in a single risk factor, by preserving a good speed up of the algorithm.
\item {[$\risk{3}$]} \label{itm:B3} We consider $M-1$ risk factors (only for $M>1$, in case $M=1$ we revert to $\risk{1}$): the values of the stock price $S_{T_{i-M+2}},\ldots,S_{T_{i}}$. We do not consider $S_{T_i-M+1}$ because the continuation value does not depend on it, since it does not contribute to the future averages. This should be the target choice if we want to consider all the relevant risk factors without introducing any approximation.
\item {[$\risk{4}$]} \label{itm:B4} We consider $n$ risk factors: the value of the stock price from the first available date up to time $T_i$. This choice considers any risk factor, not only what is required. We introduce this choice for the risk factors to understand how the LSMC method can be deceived by the presence of too many risk factors.
\end{itemize}

In Table~\ref{tab:polydim} we report the number of basis functions calculated for different values of $M$ in the worst case (when the regression date is just before the maturity date).

\begin{table}
 \begin{center}
  \pgfplotstabletypeset[
      col sep=&, row sep=\\, set thousands separator={},
      every head row/.style={before row=\toprule,after row=\midrule},
      every last row/.style={after row=\bottomrule},
      columns/rf/.style={column type=c|,column name={{\boldmath $\rho$}}},
      columns/M2/.style={column name={$\mathbf{2}$}},
      columns/M3/.style={column name={$\mathbf{3}$}},
      columns/M4/.style={column name={$\mathbf{4}$}},
      columns/M5/.style={column name={$\mathbf{5}$}},
      columns/M10/.style={column name={$\mathbf{10}$}},
      columns/M20/.style={column name={$\mathbf{20}$}},
      columns/M30/.style={column name={$\mathbf{30}$}}
  ]{
  rf &   M2 &   M3 &   M4 &   M5 &  M10 &  M20 &  M30 \\
   1 &    3 &    3 &    3 &    3 &    3 &    3 &    3 \\
   2 &    6 &    6 &    6 &    6 &    6 &    6 &    6 \\
   3 &    3 &    6 &   10 &   15 &   55 &  210 &  465 \\
   4 & 1275 & 1275 & 1275 & 1275 & 1275 & 1275 & 1275 \\}
 \end{center}
 \caption[Number of polynomial basis functions]{Number of polynomial basis functions of second degree for different values of $M$ in the worst case (when the regression date is just before the maturity date).}
 \label{tab:polydim}
\end{table}

We recall that for the R-RNN and signature cases all the observations of the underlying asset are included in the basis functions. Loosely speaking we could say that we use the risk-factor set $\risk{4}$ for such cases.

Once we have selected a specific set of risk factors we define the basis functions in the following way
\begin{equation}
 \varphi_b(S_{T_1}, \ldots, S_{T_i}) := f_b(X^\rho_1, \ldots, X^\rho_{F^\rho_i})
\end{equation}
where $f_b$ is either a polynomial, a random neural network or an element of the truncated signature. The random variables $\{X^\rho_1,\ldots,X^\rho_{F^\rho_i}\}$ are defined as
\begin{equation}
 X^\rho_j := \frac{Z^\rho_j - {\bar Z}^\rho_j}{\std{Z^\rho_j}}
 \;,\quad
 Z^\rho_j := \log \risk{\rho}_j
\end{equation}
with $1\le j\le F^\rho_i$, where $\risk{\rho}_j$ is the $j$-th risk factor of the set $\risk{\rho}$, which, in turn, depends on the realization of the price process $S_t$. Moreover, we denote the empirical mean with a bar over a random variable and the empirical standard deviation with the operator $\std{\cdot}$. The normalization of the risk factor is introduced to improve the numerical stability of the linear regression performed in the LSMC method. In the following sections we use the term ``basis function'' also to refer to $f_b$, and not only to $\varphi_b$.

Finally, in order to significantly increase the efficiency of the algorithms and decrease the computational time, we include in the various regressions only paths for which the options are in the money; see~\textcite{Longstaff2001}.

\subsection{Configuration of random networks}
\label{sec:nnsetting}

We continue the presentation of the algorithms by discussing the architectures of the randomized neural networks, in particular the type of activation function and the distribution from which the parameters are sampled. The number of hidden layers and nodes per layers for all the algorithms will be discussed later on.

We use the leaky ReLU activation function for the R-FFNN and the tanh for the R-RNN; see, also, the choice in~\textcite{herrera2021optimal}.

The parameters $(A, b)$ of the R-FFNN are sampled using a standard normal distribution with mean $0$ and standard deviation $1$, which is the standard choice. In general, $A$ and $b$ can be sampled from different distributions that are continuous and have support $\mathds{R}$. Different hyper-parameters were tested but they did not have a big influence on the results, so we kept the standard choice.

For the R-RNN we use a standard deviation of $0.0001$ for $A_x$ and $0.3$ for $A_h$. Also here different hyper-parameters were tested, and the best performing were chosen and used to present the results. In particular, these values are the values in output of a cross-validation procedure, in the sense that we test also different hyper-parameters and the best performing were chosen and used to present the results.

The parameters $\theta_{i}$ in R-FFNN and R-RNN are determined using $20\%$ of the sampled paths (training data). Given $\theta_i$, the remaining $80\%$ of the sampled paths (evaluation data) are used to compute the option price. Indeed, if the data set was not split, it might happen that the continuation value $C_{T_i}$ depends on future value of the underlying asset, and the neural network can suffer from over-fitting to the training data, by memorizing the paths, instead of learning the continuation value. More precisely, on the training data the approximation of the continuation value $C_{T_i}$ can depend on the future values of paths, since it is trained with them and therefore might remember them. However, by splitting the data into the training and the independent evaluation set, $C_{T_i}$ evaluated on the evaluation set is independent of the future values of paths of the evaluation set. Since the price is only computed on the evaluation set, this effectively prevents this ``look into the future" when computing the price. It is, however, possible that methods over-fit on the training data leading to worse results on the evaluation set.

\subsection{Configuration of signature methods}
\label{sec:sigsetting}

For the signature method, we use a truncated signature of order two, three and five. Indeed, the components of the signatures have factorial decay (see Appendix~\ref{app:signature}, Proposition~\ref{prop::factorial}), which means that it is common to choose only the first terms since these will typically be the largest. However, whilst such a truncation captures the largest components, it nevertheless loses information captured by the higher order terms.

For this reason, as explained in Subsection~\ref{sec:signature}, we use a suitable augmentation of the dataset before taking the truncated signature. More precisely, we first perform the time-augmentation (\textcite{hambly2010uniqueness}) by adding time as one of the variables, which ensures that the resulting path is uniquely determined by its signatures. Then, we apply the lead-lag transformation followed by the time-joined transformation in order to embed the time-augmented time series into a continuous path (see Subsection~\ref{sec:signature}). In this case, the resulting time series in input of the truncated signature is a $d=3$ dimensional time time series, whereas in output we have a vector of dimension $s_d(n) = (d^{n+1}-1)/(d-1)$, opportunely augmented depending on the payoff under consideration. In Subsection~\ref{subsec:sigandrs}, we will comment on the advantages (or not) of the employed techniques to augment the dataset.

In Table~\ref{tab:sigdim} we report the number of basis functions calculated for different values of $n$, the degree of the truncated signature. Although we show its basis dimensionality, note that we did not employ degree $n=4$ in our numerical experiments. We show results for dimension $d=3$, because of the application of the time-augmentation and lead-lag transformation embedding techniques.

\begin{table}
  \begin{center}
  \pgfplotstabletypeset[
      col sep=&, row sep=\\, set thousands separator={},
      every head row/.style={before row=\toprule,after row=\midrule},
      every last row/.style={after row=\bottomrule},
      columns/D2/.style={column name={$\mathbf{2}$}},
      columns/D3/.style={column name={$\mathbf{3}$}},
      columns/D4/.style={column name={$\mathbf{4}$}},
      columns/D5/.style={column name={$\mathbf{5}$}}
  ]{
    D2 &   D3 &   D4 &   D5 \\
    12 &   39 &  120 &  363 \\}
 \end{center}
  \caption[Number of signature basis functions]{Number of signature basis functions for different values of $n$, where $n$ denotes the order of the truncated signature, for path of dimension $d=3$. The results are independent of the windows length and of the risk factor set. The formula for the number of signature basis functions is given by Equation~\eqref{eq: sig_dim}.}
  \label{tab:sigdim}
\end{table}

As regards the randomized signatures, the different parameters considered in the pricing procedure are the following:
\begin{enumerate}
  \item $\varsigma$, which is the variance of the independent normal random variables used to populate the random matrices $A_i$ and biases $b_i$ of Equation~\eqref{eq:ran_sig} used in the algorithm;
  \item $k^{\star}$, which is the dimension of reservoir system, i.e~the same matrices and biases found in Equation~\eqref{eq:ran_sig};
  \item `normalization' (shortened to `norm' in the tables), which defines whether the matrices are normalized, i.e.~every entry of the random matrices $A_i$ in Equation~\eqref{eq:ran_sig} is divided by the Frobenius norm of the matrix itself.
\end{enumerate}

\section{Numerical investigations}
\label{sec:results}

This section reports and discusses the results of the numerical experiments. We investigate the performance of the LSMC method for different choices of basis functions when evaluating prices and sensitivities of Asian and look-back payoffs, the payoffs we introduced in Sections~\ref{sec:amerasian}. We add after this section a further numerical section with a complementary discussion on callable certificates to understand how our results can be extended in a more challenging setting.

The evaluation of all the algorithms is performed on the same computer, equipped with an Apple M1 Pro processor and 16 Gigabyte of RAM. Statistical uncertainties are reported at one-sigma confidence level. Computational times do not include the time for generating the stock paths (around $9$ seconds per $10^6$ paths), since it is the same for all methods. Indeed, the main interest is in the actual time needed by the algorithms to compute prices from different bases with linear regression, while paths can be generated offline and stored. Numerical procedures, unless specified otherwise, have been implemented in Python.

\subsection{Asian options}
\label{sec:asian_options}

We go on with discusses the pricing of Asian options with early termination features with both floating and fixed-strike payoff (see Equation~\eqref{eq:asian}), as well as numerical results for sensitivity computation only for the fixed-strike versions of the payoff since the floating-strike ones have trivial Delta and Gamma.

In particular, Table~\ref{tab:afs_prices} (resp.~Table~\ref{tab:axs_prices}) lists option prices obtained for different lengths $M$ of the moving window, obtained via the LSMC method with different choices of basis functions, namely polynomials, R-FFNN, R-RRN, and signatures methods, for Asian options with floating-strike (resp.~fixed-strike) payoff. Table~\ref{tab:afs_errors} (resp.~Table~\ref{tab:axs_errors}) and Table~\ref{tab:afs_times} (resp.~Table~\ref{tab:afs_times}) report the corresponding uncertainties of Monte Carlo simulations and computational times, respectively.   

\begin{table}
\begin{center}
  \pgfplotstabletypeset[
    skip rows between index={29}{31},
    string replace={0}{},
    fixed, fixed zerofill, precision=3,
    every head row/.style={before row=\toprule\multicolumn{8}{l}{\bf Floating-strike Asian option: prices}\\\toprule,after row=\midrule},
    every last row/.style={after row=\bottomrule},
    every row no 5/.style={before row=\hline},
    every row no 9/.style={before row=\hline},
    every row no 13/.style={before row=\hline},
    every row no 17/.style={before row=\hline},
    every row no 21/.style={before row=\hline},
    every row no 25/.style={before row=\hline},
    every row no 29/.style={before row=\hline},
    every row no 32/.style={before row=\hline},
    columns={model,afs2p,afs3p,afs4p,afs5p,afs10p,afs20p,afs30p},
    columns/model/.style={string type,column type=l|,column name={\bf Model}},
    columns/afs2p/.style={column name={$\mathbf{2}$}},
    columns/afs3p/.style={column name={$\mathbf{3}$}},
    columns/afs4p/.style={column name={$\mathbf{4}$}},
    columns/afs5p/.style={column name={$\mathbf{5}$}},
    columns/afs10p/.style={column name={$\mathbf{10}$}},
    columns/afs20p/.style={column name={$\mathbf{20}$}},
    columns/afs30p/.style={column name={$\mathbf{30}$}},
  ]{\afsdat}
\end{center}
\caption[Prices of floating-strike American-style Asian options]{Prices of floating-strike American-style Asian options with maturity $T=0.2$ and time steps $N=50$ for different lengths of the moving window. Comparison between different models (see text). Benchmark models: GPR-GHQ and binomial Markov chain of~\textcite{Goudenege2022}, \textcite{bernhart2011finite}, \textcite{lelong2019pricing}. LSMC models: polynomial bases, randomized neural networks (R-FFNN and R-RNN), signature and randomized signature (RandSig) based methods.}
\label{tab:afs_prices}
\end{table}

\begin{table}
\begin{center}
  \pgfplotstabletypeset[
    skip rows between index={29}{31},
    string replace={0}{},
    fixed, fixed zerofill, precision=0, set thousands separator={},
    every head row/.style={before row=\toprule\multicolumn{8}{l}{\bf Floating-strike Asian option: computational times}\\\toprule,after row=\midrule},
    every last row/.style={after row=\bottomrule},
    every row no 5/.style={before row=\hline},
    every row no 9/.style={before row=\hline},
    every row no 13/.style={before row=\hline},
    every row no 17/.style={before row=\hline},
    every row no 21/.style={before row=\hline},
    every row no 25/.style={before row=\hline},
    every row no 29/.style={before row=\hline},
    every row no 32/.style={before row=\hline},
    columns={model,afs2t,afs3t,afs4t,afs5t,afs10t,afs20t,afs30t},
    columns/model/.style={string type,column type=l|,column name={\bf Model}},
    columns/afs2t/.style={column name={$\mathbf{2}$}},
    columns/afs3t/.style={column name={$\mathbf{3}$}},
    columns/afs4t/.style={column name={$\mathbf{4}$}},
    columns/afs5t/.style={column name={$\mathbf{5}$}},
    columns/afs10t/.style={column name={$\mathbf{10}$}},
    columns/afs20t/.style={column name={$\mathbf{20}$}},
    columns/afs30t/.style={column name={$\mathbf{30}$}},
  ]{\afsdat}
\end{center}
\caption[Computational time for floating-strike American-style Asian options]{Algorithm computational time (in seconds) to price floating-strike American-style Asian options with maturity $T=0.2$ and time steps $N=50$ for different lengths of the moving window. Comparison between different models (see text). Benchmark models: GPR-GHQ and binomial Markov chain of~\textcite{Goudenege2022}, \textcite{bernhart2011finite}, \textcite{lelong2019pricing}. LSMC models: polynomial bases, randomized neural networks (R-FFNN and R-RNN), signature and randomized signature (RandSig) based methods.}
\label{tab:afs_times}
\end{table}

\begin{table}
\begin{center}
  \pgfplotstabletypeset[
    skip rows between index={24}{26},
    string replace={0}{},
    fixed, fixed zerofill, precision=3,
    every head row/.style={before row=\toprule\multicolumn{8}{l}{\bf Fixed-strike Asian option: prices}\\\toprule,after row=\midrule},
    every last row/.style={after row=\bottomrule},
    every row no 4/.style={before row=\hline},
    every row no 8/.style={before row=\hline},
    every row no 12/.style={before row=\hline},
    every row no 16/.style={before row=\hline},
    every row no 20/.style={before row=\hline},
    every row no 24/.style={before row=\hline},
    every row no 28/.style={before row=\hline},
    columns={model,axs2p,axs3p,axs4p,axs5p,axs10p,axs20p,axs30p},
    columns/model/.style={string type,column type=l|,column name={\bf Model}},
    columns/axs2p/.style={column name={$\mathbf{2}$}},
    columns/axs3p/.style={column name={$\mathbf{3}$}},
    columns/axs4p/.style={column name={$\mathbf{4}$}},
    columns/axs5p/.style={column name={$\mathbf{5}$}},
    columns/axs10p/.style={column name={$\mathbf{10}$}},
    columns/axs20p/.style={column name={$\mathbf{20}$}},
    columns/axs30p/.style={column name={$\mathbf{30}$}},
  ]{\axsdat}
\end{center}
\caption[Prices of fixed-strike American-style Asian options]{Prices of fixed-strike American-style Asian options with strike $K=100$, maturity $T=0.2$ and time steps $N=50$ for different lengths of the moving window. Comparison between different models (see text). LSMC models: polynomial bases, randomized neural networks (R-FFNN and R-RNN), signature and randomized signature (RandSig) based methods.}
\label{tab:axs_prices}
\end{table}
  
\begin{table}
\begin{center}
  \pgfplotstabletypeset[
    skip rows between index={24}{26},
    string replace={0}{},
    fixed, fixed zerofill, precision=0, set thousands separator={},
    every head row/.style={before row=\toprule\multicolumn{8}{l}{\bf Fixed-strike Asian option: computational times}\\\toprule,after row=\midrule},
    every last row/.style={after row=\bottomrule},
    every row no 4/.style={before row=\hline},
    every row no 8/.style={before row=\hline},
    every row no 12/.style={before row=\hline},
    every row no 16/.style={before row=\hline},
    every row no 20/.style={before row=\hline},
    every row no 24/.style={before row=\hline},
    every row no 28/.style={before row=\hline},
    columns={model,axs2t,axs3t,axs4t,axs5t,axs10t,axs20t,axs30t},
    columns/model/.style={string type,column type=l|,column name={\bf Model}},
    columns/axs2t/.style={column name={$\mathbf{2}$}},
    columns/axs3t/.style={column name={$\mathbf{3}$}},
    columns/axs4t/.style={column name={$\mathbf{4}$}},
    columns/axs5t/.style={column name={$\mathbf{5}$}},
    columns/axs10t/.style={column name={$\mathbf{10}$}},
    columns/axs20t/.style={column name={$\mathbf{20}$}},
    columns/axs30t/.style={column name={$\mathbf{30}$}},
  ]{\axsdat}
\end{center}
\caption[Computational time for fixed-strike American-style Asian options]{Algorithm computational time (in seconds) to price fixed-strike American-style Asian options with strike $K=100$, maturity $T=0.2$ and time steps $N=50$ for different lengths of the moving window. Comparison between different models (see text). LSMC models: polynomial bases, randomized neural networks (R-FFNN and R-RNN), signature and randomized signature (RandSig) based methods.}
\label{tab:axs_times}
\end{table}
  
In all the computations the option maturity is $T=0.2$ and the number of time steps is $N=50$. The LSMC results are obtained by batches of simulations with overall $8\cdot 10^5$ paths to train the regression and $3.2\cdot 10^6$ to calculate the price.

We continue the analysis by discussing the benchmark models, then we go on with the LSMC algorithm with polynomials, random networks and signatures as basis functions. 
The numerical results (prices and computational times) can be found for floating-strike options in Tables~\ref{tab:afs_prices} and~\ref{tab:afs_times}, while for fixed-strike options in Tables~\ref{tab:axs_prices} and~\ref{tab:axs_times}. In Appendix~\ref{app:numerical_data} we display also the corresponding statistical uncertainties in Tables~\ref{tab:afs_errors} and~\ref{tab:axs_errors}.

\subsubsection{Benchmarks}

We search the literature for benchmark models. We found three papers focusing on fixed-strike Asian options with observations performed in a moving window: \textcite{bernhart2011finite}, \textcite{lelong2019pricing} and \textcite{Goudenege2022}. In the first five rows in Tables~\ref{tab:afs_prices}, \ref{tab:afs_times} and~\ref{tab:afs_errors} we list their results.

In~\textcite{Goudenege2022} the authors present an efficient method for pricing American-style moving average options based on Gaussian Process Regression and Gauss-Hermite quadrature, thus named GPR-GHQ method (GPR in our tables, for short); $P$ denotes the number of discrete paths for the price process to compute the GPR, whereas $Q$ the number of points employed in the GHQ. Moreover, by exploiting the positive homogeneity of the continuation value, they develop a binomial Markov chain to deal efficiently with medium-long windows (Binomial Chain in our tables). We review both methods in Appendix~\ref{app:Zanette}. In their numerical investigation, they compare their two methodologies primarily with LSMC. We refer the readers to the authors' original papers for a discussion of the other two contributions.

In detail, among the benchmark models we present GPR-GPQ by~\textcite{Goudenege2022} with two different choices for parameters $P$ and $Q$, the method of~\textcite{bernhart2011finite} and the one of~\textcite{lelong2019pricing}. In our analysis we compare the LSMC method with the benchmark models and we discuss the impact of selecting different sets of basis functions.

\subsubsection{Polynomials}

The simplest choice of basis functions are polynomials. \textcite{Longstaff2001} proposed to use the first three weighted Laguerre polynomials, the first three Hermite polynomials, three trigonometric functions, and simple powers of the underline as basis function obtaining results that are virtually identical to each others. In particular, classical polynomial basis functions up to the second order are the the easiest way to include coupling terms in the basis. The results obtained with the classical polynomials up to degree two were better than with the weighted Laguerre polynomials, therefore we only present these results in our Tables~\ref{tab:afs_prices} and~\ref{tab:axs_prices} (in the Tables \textit{deg} refers to the maximum degree of the polynomial). Considerations on the results are consistent for the two types of payoffs.

Empirically, the risk set $\risk{2}$ is preferred to the risk set $R^3$, although the latter should contain the best set of information to compute the option prices. In particular, it seems that the choice of $\risk{3}$ loses effectiveness in case of large values of $M$. In the case of floating-strike Asian options, the risk set $\risk{2}$ can beat in some cases also the results obtained with our benchmark model GPR-GPQ. We notice also, by looking at Table~\ref{tab:afs_times}, that the LSMC with risk set $\risk{2}$ is to be preferred also in term of calculation speed, especially when compared with the GPR-GPQ method with $P=1000$ and $Q=16$. We notice that \textcite{Goudenege2022} employ $\risk{3}$ in their work, thus leading to the wrong conclusion that the continuation value provided by GPR-GPQ is more accurate than that provided by the LSMC method.

In the $\risk{4}$ case the ability to approximate the continuation value, especially for large values of $M$ is limited by the use of a polynomial of degree two, a limitation imposed by the large size of the problem ($1275$ in this case).

\subsubsection{Randomized neural networks}

The choice of polynomials as basis functions may be too simple to catch the non-linear behavior of the continuation value. Thus, we investigate the possibility to replace them by means of a neural network.

We start by discussing R-FFNN basis functions. In order to compare them with polynomials, for a fixed set of risk factors (or features if we wish to use a machine learning terminology), we only change the number of nodes per layer. Indeed, similarly to what found in~\textcite{herrera2021optimal}, we observed that one hidden layer was sufficient to have a good accuracy (an increase of the number of the hidden layers did not leas to better accuracy). Tables~\ref{tab:afs_prices} and~\ref{tab:axs_prices} reports the results.

Quite remarkably, also for the R-FFNN, the results obtained by means of $\risk{2}$ are slightly higher than those obtained through other risk-factors sets, which indicates that $\risk{2}$ is more effective than the other bases; this fact is confirmed for every employed value of the hidden size. By comparing the results with polynomial basis functions, we can see that a hidden size equals to $40$ is necessary in order to reach the same pricing accuracy. For such value of the hidden size, the computational times of the two pricing techniques are comparable.

Tables~\ref{tab:afs_prices} and~\ref{tab:axs_prices} display the results for R-RNNs. We formulate the following observations. First, the hidden size does not seem to affect the results. Second, the computational times grow linearly with the hidden size, and there is a tangible reduction in time with respect to the others methods, when applied to risk set $\risk{4}$. However, the results obtained with the R-FFNN and polynomials with risk-factor set $\risk{2}$ are higher than those obtained with R-RNNs.

\subsubsection{Signature and randomized signature methods}
\label{subsec:sigandrs}

Tables~\ref{tab:afs_prices} and~\ref{tab:axs_prices} display the results. For signatures methods there are relevant gains in the effectiveness if we truncate the signature at order five instead of order two or three. However, similarly to what happens with the R-RNNs, the results obtained with the R-FFNN and polynomials with risk-factor set $\risk{2}$ are higher.

Before proceeding, the following remark is in order. As said in Section~\ref{sec:signature}, one may wonder if the lead-lag transformation is strictly necessary since it duplicates the (spatial) dimension of the time-series. We thus compute prices of fixed-strike American-style options in the same setting of Table~\ref{tab:afs_prices}, with $n=3$ and without the lead-lag transformation. As you can see the prices obtained with the lead-lag transformation are statistically higher than the ones without such embedding; however, the computational time is much higher.

In the same Tables~\ref{tab:afs_prices} and~\ref{tab:axs_prices} we also include randomized signatures, for which a light introduction is provided in Section~\ref{sec:rand_signature}. We tried this algorithm only for floating-strike and fixed-strike American-style Asian options.

Results show a promising performance for floating-strike options for short-time windows, with respect to standard signature methods, where one can observe that the advantage of using this technique deteriorates while increasing the window length. For the chosen parameters, its prominence decreases for window lengths higher than $5$. Results are less clear, but quite similar, for fixed-strike options, although the difference with the polynomial and other methods is less evident. In both cases, it is possible to see that normalization can help the stability of the method and improves pricing results. Computational times are quite high if compared with polynomials on $\risk{2}$ or low-dimensional neural networks, but much lower when compared with standard signatures. For this reason and the higher calculated derivative prices, randomized signatures are preferred over standard signatures.

\subsubsection{Sensitivity computation}
\label{subsec:sensitivity_asian}

The discussion in Section~\ref{sec:sensitivity} show us that finite-difference and Chebyshev methods can be effectively used to calculate sensitivities, although the former method suffers biases, while the regression method seems less reliable. In this section we repeat the analysis for fixed-strike Asian options, by limiting ourself to the first two methods. In Table~\ref{tab:gamma_axs} we can see the results, which are qualitatively similar to the ones of the American put case. In Appendix~\ref{app:sensitivity} we extend the analysis.

\begin{table}
\begin{center}
  \pgfplotstabletypeset[
    string replace={0}{},
    fixed, fixed zerofill, precision=4, set thousands separator={},
    every head row/.style={after row=\midrule,before row=
      \toprule
      \multicolumn{9}{l}{\bf Fixed-Strike Asian option: Gamma} \\
      \midrule
      & \multicolumn{4}{c|}{$\bm{S_0=85}$}
      & \multicolumn{4}{c}{$\bm{S_0=115}$} \smallskip\\
      \midrule
      & \multicolumn{2}{c|}{\small \bf Finite Difference}
      & \multicolumn{2}{c|}{\small \bf Chebyshev}
      & \multicolumn{2}{c|}{\small \bf Finite Difference}
      & \multicolumn{2}{c}{\small \bf Chebyshev} \smallskip\\
      },
    every last row/.style={after row=\bottomrule},
    columns={npath,ye085,fe085,yc085,fc085,ye115,fe115,yc115,fc115},
    columns/npath/.style={sci, precision=1, column name={\small \bf MC Path}},
    columns/ye085/.style={column name={\small \bf Polyn.}, 
    column type/.add={|}{}},
    columns/fe085/.style={column name={\small \bf FFNN}},
    columns/yc085/.style={column name={\small \bf Polyn.}, 
    column type/.add={|}{}},
    columns/fc085/.style={column name={\small \bf FFNN}},
    columns/ye115/.style={column name={\small \bf Polyn.}, 
    column type/.add={|}{}},
    columns/fe115/.style={column name={\small \bf FFNN}},
    columns/yc115/.style={column name={\small \bf Polyn.}, 
    column type/.add={|}{}},
    columns/fc115/.style={column name={\small \bf FFNN}},
  ]{\sensmcaxs}
\end{center}
\caption[Gamma for a fixed-strike Asian option]{Gamma for a fixed-strike Asian option with moving window $M=2$, maturity $T=0.2$, strike $K=100$ and time steps $N=50$. The first column is the total number of simulation paths used by each algorithm. Then, there are two groups each of four columns for different spot prices $S_0$. In each group the first two columns is the result of the finite-difference method ($\epsilon=1/8$) with polynomials or R-FFNN basis functions, the last two columns of the Chebyshev method ($\epsilon=5/8$).}
\label{tab:gamma_axs}
\end{table}

The previous results lead us to confirm the use of the Chebyshev method for sensitivity computations. Figure~\ref{fig:axs_sensitivity} shows the results for Delta and Gamma sensitivities. We calculate option sensitivities for different lengths $M$ of the moving-window by means of the LSMC method with polynomials basis functions with risk factor set $\risk{3}$. In all the computations the option maturity is $T=0.2$, number of time steps is $N=50$, the number of paths is $2.5\cdot 10^5$ both to train the regression and to calculate the price. The sensitivities are calculated by means of the Chebyshev interpolation with an adaptive interpolation interval of ten percent of the spot price according to the algorithm described in~\textcite{Maran2020}.

We can see that the Gamma function is not discontinuous, as in the American put case described in Section~\ref{sec:sensitivity}, since in the present case we wait for $M-1$ days to start the early-termination period, as described in Section~\ref{sec:amerasian}. However, its value abruptly increases spot price around $80\%$ to decrease back to zero at higher spot prices. We notice minor noisy area in calculating the Gamma sensitivity, while Delta is always very smooth.

\begin{figure}
  \ifx\RemoveFigures\undefined
    \begin{center}
    \scalebox{0.9}{%
    \begin{tikzpicture}
    \begin{axis}[xmin=1,xmax=30,
              ytick={0.75,1,1.25},
              yticklabels={75,100,125},
              grid=major]
    \addplot3 [color=mydarkblue,thick,smooth] table [z=d2,y=money,x expr=2] from \sensaxs;
    \addplot3 [color=mydarkblue,thick,smooth] table [z=d3,y=money,x expr=3] from \sensaxs;
    \addplot3 [color=mydarkblue,thick,smooth] table [z=d4,y=money,x expr=4] from \sensaxs;
    \addplot3 [color=mydarkblue,thick,smooth] table [z=d5,y=money,x expr=5] from \sensaxs;
    \addplot3 [color=mydarkblue,thick,smooth] table [z=d10,y=money,x expr=10] from \sensaxs;
    \addplot3 [color=mydarkblue,thick,smooth] table [z=d20,y=money,x expr=20] from \sensaxs;
    \addplot3 [color=mydarkblue,thick,smooth] table [z=d30,y=money,x expr=30] from \sensaxs;
    \end{axis}
    \end{tikzpicture}
    \begin{tikzpicture}
    \begin{axis}[xmin=1,xmax=30,
              ytick={0.75,1,1.25},
              yticklabels={75,100,125},
              grid=major]
    \addplot3 [color=mydarkblue,thick,smooth] table [z=g2,y=money,x expr=2] from \sensaxs;
    \addplot3 [color=mydarkblue,thick,smooth] table [z=g3,y=money,x expr=3] from \sensaxs;
    \addplot3 [color=mydarkblue,thick,smooth] table [z=g4,y=money,x expr=4] from \sensaxs;
    \addplot3 [color=mydarkblue,thick,smooth] table [z=g5,y=money,x expr=5] from \sensaxs;
    \addplot3 [color=mydarkblue,thick,smooth] table [z=g10,y=money,x expr=10] from \sensaxs;
    \addplot3 [color=mydarkblue,thick,smooth] table [z=g20,y=money,x expr=20] from \sensaxs;
    \addplot3 [color=mydarkblue,thick,smooth] table [z=g30,y=money,x expr=30] from \sensaxs;
    \end{axis}
    \end{tikzpicture}}%
  \end{center}
  \fi
\caption[Delta and Gamma of fixed-strike American-style Asian options]{Delta (left panel) and Gamma (right panel) of fixed-strike American-style Asian options with strike $K=100$, maturity $T=0.2$ and time steps $N=50$ for different lengths of the moving window. LSMC model with polynomial base with risk-factor set $\risk{3}$. On the left axis is reported the window length $M$, while on the right axis the spot price.}
\label{fig:axs_sensitivity}
\end{figure}

\subsection{Look-back options}
\label{sec:lookback_num}

We repeat the analysis done in the previous Subsections~\ref{sec:asian_options} on Asian payoffs in the case of floating-strike and fixed-strike look-back ones. We notice that in this case we replace the arithmetic moving average computed over the interval $[T_{i-M+1}, T_{i}]$ with the re-scaled minimum or maxima of the asset price when computing the risk factor $\risk{2}$.

Here, we do not have a comparison with benchmark models in the literature, but we can still compare the results of our LSMC approach based on different choices of the basis functions.

Tables~\ref{tab:lfs_prices} and~\ref{tab:lxs_prices} display the results, while uncertainties of Monte Carlo simulations are reported in Appendix~\ref{app:numerical_data} in Tables~\ref{tab:lfs_errors} and~\ref{tab:lxs_errors}. In Tables~\ref{tab:lfs_times} and~\ref{tab:lxs_times} are listed the computational times.

First, we can observe that, as for the Asian options, the empirical qualitative results for the fixed-strike payoff and for the floating-strike payoff are very similar. However, in this case, for polynomials results obtained with the risk factor set $\risk{3}$ are higher than those of $\risk{2}$, especially for large value of $M$. For R-FFNN, instead, the risk factor set $\risk{2}$ is preferred to the risk set $\risk{3}$; again, a hidden size equals to 40 is necessary in order to reach the same pricing accuracy of polynomials. The result obtained with R-RNNs are smaller than those obtained with the R-FFNN and polynomials. Finally, for the signature methods an order equal to 5 is necessary to achieve the same pricing accuracy of polynomials, at a cost of a very high computational time.

\begin{table}
 \begin{center}
  \pgfplotstabletypeset[
    skip rows between index={24}{26},
    string replace={0}{},
    fixed, fixed zerofill, precision=3,
    every head row/.style={before row=\toprule\multicolumn{8}{l}{\bf Floating-strike look-back option: prices}\\\toprule,after row=\midrule},
    every last row/.style={after row=\bottomrule},
    every row no 4/.style={before row=\hline},
    every row no 8/.style={before row=\hline},
    every row no 12/.style={before row=\hline},
    every row no 16/.style={before row=\hline},
    every row no 20/.style={before row=\hline},
    every row no 24/.style={before row=\hline},
    columns={model,lfs2p,lfs3p,lfs4p,lfs5p,lfs10p,lfs20p,lfs30p},
    columns/model/.style={string type,column type=l|,column name={\bf Model}},
    columns/lfs2p/.style={column name={$\mathbf{2}$}},
    columns/lfs3p/.style={column name={$\mathbf{3}$}},
    columns/lfs4p/.style={column name={$\mathbf{4}$}},
    columns/lfs5p/.style={column name={$\mathbf{5}$}},
    columns/lfs10p/.style={column name={$\mathbf{10}$}},
    columns/lfs20p/.style={column name={$\mathbf{20}$}},
    columns/lfs30p/.style={column name={$\mathbf{30}$}},
  ]{\lfsdat}
 \end{center}
 \caption[Prices of floating-strike American-style look-back options]{Prices of floating-strike American-style look-back options with maturity $T=0.2$ and time steps $N=50$ for different lengths of the moving window. Comparison between different models (see text). LSMC models: polynomial bases, randomized neural networks (R-FFNN and R-RNN), signature based methods.}
 \label{tab:lfs_prices}
\end{table}

\begin{table}
 \begin{center}
  \pgfplotstabletypeset[
    skip rows between index={24}{26},
    string replace={0}{},
    fixed, fixed zerofill, precision=0, set thousands separator={},
    every head row/.style={before row=\toprule\multicolumn{8}{l}{\bf Floating-strike look-back option: computational times}\\\toprule,after row=\midrule},
    every last row/.style={after row=\bottomrule},
    every row no 4/.style={before row=\hline},
    every row no 8/.style={before row=\hline},
    every row no 12/.style={before row=\hline},
    every row no 16/.style={before row=\hline},
    every row no 20/.style={before row=\hline},
    every row no 24/.style={before row=\hline},
    columns={model,lfs2t,lfs3t,lfs4t,lfs5t,lfs10t,lfs20t,lfs30t},
    columns/model/.style={string type,column type=l|,column name={\bf Model}},
    columns/lfs2t/.style={column name={$\mathbf{2}$}},
    columns/lfs3t/.style={column name={$\mathbf{3}$}},
    columns/lfs4t/.style={column name={$\mathbf{4}$}},
    columns/lfs5t/.style={column name={$\mathbf{5}$}},
    columns/lfs10t/.style={column name={$\mathbf{10}$}},
    columns/lfs20t/.style={column name={$\mathbf{20}$}},
    columns/lfs30t/.style={column name={$\mathbf{30}$}},
  ]{\lfsdat}
 \end{center}
 \caption[Computational time for floating-strike American-style look-back options]{Algorithm computational time (in seconds) to price floating-strike American-style look-back options with maturity $T=0.2$ and time steps $N=50$ for different lengths of the moving window. Comparison between different models (see text). LSMC models: polynomial bases, randomized neural networks (R-FFNN and R-RNN), signature based methods.}
 \label{tab:lfs_times}
\end{table}

\begin{table}
 \begin{center}
  \pgfplotstabletypeset[
    skip rows between index={24}{26},
    string replace={0}{},
    fixed, fixed zerofill, precision=3,
    every head row/.style={before row=\toprule\multicolumn{8}{l}{\bf Fixed-strike look-back option: prices}\\\toprule,after row=\midrule},
    every last row/.style={after row=\bottomrule},
    every row no 4/.style={before row=\hline},
    every row no 8/.style={before row=\hline},
    every row no 12/.style={before row=\hline},
    every row no 16/.style={before row=\hline},
    every row no 20/.style={before row=\hline},
    every row no 24/.style={before row=\hline},
    columns={model,lxs2p,lxs3p,lxs4p,lxs5p,lxs10p,lxs20p,lxs30p},
    columns/model/.style={string type,column type=l|,column name={\bf Model}},
    columns/lxs2p/.style={column name={$\mathbf{2}$}},
    columns/lxs3p/.style={column name={$\mathbf{3}$}},
    columns/lxs4p/.style={column name={$\mathbf{4}$}},
    columns/lxs5p/.style={column name={$\mathbf{5}$}},
    columns/lxs10p/.style={column name={$\mathbf{10}$}},
    columns/lxs20p/.style={column name={$\mathbf{20}$}},
    columns/lxs30p/.style={column name={$\mathbf{30}$}},
  ]{\lxsdat}
 \end{center}
 \caption[Prices of fixed-strike American-style look-back options]{Prices of fixed-strike American-style look-back options with strike $K=100$, maturity $T=0.2$ and time steps $N=50$ for different lengths of the moving window. Comparison between different models (see text). LSMC models: polynomial bases, randomized neural networks (R-FFNN and R-RNN), signature based methods.}
 \label{tab:lxs_prices}
\end{table}

\begin{table}
 \begin{center}
  \vspace{2.5mm}
  \pgfplotstabletypeset[
    skip rows between index={24}{26},
    string replace={0}{},
    fixed, fixed zerofill, precision=0, set thousands separator={},
    every head row/.style={before row=\toprule\multicolumn{8}{l}{\bf Fixed-strike look-back option: computational times}\\\toprule,after row=\midrule},
    every last row/.style={after row=\bottomrule},
    every row no 4/.style={before row=\hline},
    every row no 8/.style={before row=\hline},
    every row no 12/.style={before row=\hline},
    every row no 16/.style={before row=\hline},
    every row no 20/.style={before row=\hline},
    every row no 24/.style={before row=\hline},
    columns={model,lxs2t,lxs3t,lxs4t,lxs5t,lxs10t,lxs20t,lxs30t},
    columns/model/.style={string type,column type=l|,column name={\bf Model}},
    columns/lxs2t/.style={column name={$\mathbf{2}$}},
    columns/lxs3t/.style={column name={$\mathbf{3}$}},
    columns/lxs4t/.style={column name={$\mathbf{4}$}},
    columns/lxs5t/.style={column name={$\mathbf{5}$}},
    columns/lxs10t/.style={column name={$\mathbf{10}$}},
    columns/lxs20t/.style={column name={$\mathbf{20}$}},
    columns/lxs30t/.style={column name={$\mathbf{30}$}},
  ]{\lxsdat}
 \end{center}
 \caption[Computational time for fixed-strike American-style look-back options]{Algorithm computational time (in seconds) to price fixed-strike American-style look-back options with strike $K=100$, maturity $T=0.2$ and time steps $N=50$ for different lengths of the moving window. Comparison between different models (see text). LSMC models: polynomial bases, randomized neural networks (R-FFNN and R-RNN), signature based methods.}
\label{tab:lxs_times}
\end{table}

\subsubsection{Sensitivity computation}
\label{subsec:sensitivity_lookback}

Figure~\ref{fig:lxs_sensitivity} shows the results for Delta and Gamma sensitivities for the fixed-strike version of the payoff. We calculate option sensitivities for different lengths $M$ of the moving-window by means of the LSMC method with polynomials basis functions with risk factor set $\risk{3}$ with the same settings used for the fixed-strike Asian payoff.

We can see that the Gamma function is continuous, as in the fixed-strike Asian option case, see Section~\ref{subsec:sensitivity_asian}. As in such case we notice minor noisy area in calculating the Gamma sensitivity, while Delta is always very smooth.

\begin{figure}
\ifx\RemoveFigures\undefined
  \begin{center}
  \scalebox{0.9}{%
    \begin{tikzpicture}
      \begin{axis}[xmin=1,xmax=30,
                  ytick={0.75,1,1.25},
                  yticklabels={75,100,125},
                  grid=major]
      \addplot3 [color=mydarkblue,thick,smooth] table [z=d2,y=money,x expr=2] from \senslxs;
      \addplot3 [color=mydarkblue,thick,smooth] table [z=d3,y=money,x expr=3] from \senslxs;
      \addplot3 [color=mydarkblue,thick,smooth] table [z=d4,y=money,x expr=4] from \senslxs;
      \addplot3 [color=mydarkblue,thick,smooth] table [z=d5,y=money,x expr=5] from \senslxs;
      \addplot3 [color=mydarkblue,thick,smooth] table [z=d10,y=money,x expr=10] from \senslxs;
      \addplot3 [color=mydarkblue,thick,smooth] table [z=d20,y=money,x expr=20] from \senslxs;
      \addplot3 [color=mydarkblue,thick,smooth] table [z=d30,y=money,x expr=30] from \senslxs;
      \end{axis}
      \end{tikzpicture}
      \begin{tikzpicture}
      \begin{axis}[xmin=1,xmax=30,
                  ytick={0.75,1,1.25},
                  yticklabels={75,100,125},
                  grid=major]
      \addplot3 [color=mydarkblue,thick,smooth] table [z=g2,y=money,x expr=2] from \senslxs;
      \addplot3 [color=mydarkblue,thick,smooth] table [z=g3,y=money,x expr=3] from \senslxs;
      \addplot3 [color=mydarkblue,thick,smooth] table [z=g4,y=money,x expr=4] from \senslxs;
      \addplot3 [color=mydarkblue,thick,smooth] table [z=g5,y=money,x expr=5] from \senslxs;
      \addplot3 [color=mydarkblue,thick,smooth] table [z=g10,y=money,x expr=10] from \senslxs;
      \addplot3 [color=mydarkblue,thick,smooth] table [z=g20,y=money,x expr=20] from \senslxs;
      \addplot3 [color=mydarkblue,thick,smooth] table [z=g30,y=money,x expr=30] from \senslxs;
      \end{axis}
    \end{tikzpicture}}%
  \end{center}
\fi
\caption[Delta and Gamma of fixed-strike American-style look-back options]{Delta (left panel) and Gamma (right panel) of fixed-strike American-style look-back options with strike $K=100$, maturity $T=0.2$ and time steps $N=50$ for different lengths of the moving window. 
LSMC model with polynomial base with risk-factor set $\risk{3}$. 
On the left axis is reported the window length $M$, while on the right axis the spot price.}
\label{fig:lxs_sensitivity}
\end{figure}

\section{Callable certificates}
\label{sec:callable}

In the previous numerical section we have seen that polynomial basis functions or random networks, in particular the R-FFNN architecture, seem the best choices whenever we can prune the input features of unnecessary information, as we do when we choose the risk factor $\risk{2}$.

However, we must exploit the specific form tof the payoff to devise the right risk-factor set. In general, if we deal with more complex situations, we could not introduce such simplifications. In order to understand the challenge of this cases we analyze a class of investment products actively traded in the market: callable certificates.

\subsection{Product description}
\label{sec:callable-description}

Certificates are structured investment products which usually contains optional features automatically triggered based on a specific event. A common type of equity certificate is the auto-callable product. These products are triggered on predetermined observation dates if an underlying asset or reference portfolio reaches or surpasses a barrier. Upon termination, the investor receives the principal investment amount plus a coupon. If the product is not terminated early, the investor receives the principal amount, or a portion thereof, along with an optional payoff on the maturity date. This feature is frequently offered in low-yield markets, providing the investor with the potential for an above-market yield, albeit with the risk of not receiving a coupon. We refer the readers to~\textcite{Deng2016,Alm2013,Farkas2022} and references therein for a description of this product.

In this study, we analyze equity certificates that feature an early-termination clause instead of an auto-callable feature. This allows the issuer to terminate the product on predetermined observation dates at their discretion, making them ``callable certificates.'' This enables the issuer to have greater control over the marking of their funding benefits, while the investor still has the opportunity to re-enter a high-yield structure on termination dates. There is a limited literature on these new products. We refer the readers to~\textcite{Agrawal2022} and references therein for further details. 

In this section we present the numerical results for ``snowball'' and ``lock-in'' callable certificates. The specifics of the payoffs and the corresponding dynamic programming problem are presented in Appendix~\ref{app:callable}. In this numerical section we assume that the underlying asset follows the dynamics described in Section~\ref{sec:dynamics}.

We perform all the numerical investigations by using a LSMC method with different choices of the basis functions. As risk factors we use at regression date $T_i$, with $1\le i\le N$ and $T_N=T$, a set formed by the observations of the underlying price processes on all the coupon dates up to the regression date.

\subsection{Snowball payoff}
\label{subsec:snowball_num}

We consider a snowball payoff with different maturities which pays quarterly a coupon if the underlying asset performance is above $100\%$. In particular we choose the following characteristics: (i) maturity of one year with coupon equal to $2.3\%$ and capital barrier at $35\%$, (ii) maturity of two years with coupon equal to $2.4\%$ and capital barrier at $30\%$, (iii) maturity of five years with coupon equal to $2.85\%$ and capital barrier at $30\%$. The characteristics of the product are chosen so that the initial price is approximately at par, when evaluated with the LSMC method with polynomial basis on the risk set described at the beginning of Section~\ref{sec:callable}.

In the first three columns of Table~\ref{tab:certificate_prices} we report the prices of the snowball certificates defined above by changing the basis functions used in the LSMC method to price the products. We can see that the prices are closely aligned, and within the Monte Carlo uncertainties, reported in appendix in Table~\ref{tab:certificate_errors}. In Table~\ref{tab:certificate_times} are listed the computational times. We can see the best choice of basis functions is R-FFNN and R-RNN (in particular for longer maturities), followed by polynomials. In particular, the last ones become unfeasible for many dates and larger degrees (in the Tables \textit{deg} refers to the maximum degree of the polynomial).

\begin{table}
 \begin{center}
  \pgfplotstabletypeset[
    string replace={0}{},
    fixed, fixed zerofill, precision=5,
    every head row/.style={before row=\toprule{\bf Certificate: prices}&\multicolumn{3}{c|}{\bf Snowball}&\multicolumn{3}{c}{\bf Lock-in}\\\toprule,after row=\midrule},
    every last row/.style={after row=\bottomrule},
    every row no 3/.style={before row=\hline},
    every row no 6/.style={before row=\hline},
    every row no 9/.style={before row=\hline},
    columns={model,sn1p,sn2p,sn5p,li1p,li2p,li5p},
    columns/model/.style={string type,column type=l|,column name={\bf Model}},
    columns/sn1p/.style={column name={\bf 1y}},
    columns/sn2p/.style={column name={\bf 2y}},
    columns/sn5p/.style={column type=c|,column name={\bf 5y}},
    columns/li1p/.style={column name={\bf 1y}},
    columns/li2p/.style={column name={\bf 2y}},
    columns/li5p/.style={column name={\bf 5y}},
  ]{\certificatedat}
 \end{center}
 \caption[Prices of snowball and lock-in certificates]{Prices of snowball and lock-in certificates with maturity of 1, 2 and 5 years (other characteristics described in the text). Comparison between different models (see text). LSMC models: polynomial bases, randomized neural networks (R-FFNN and R-RNN), signature based methods.}
 \label{tab:certificate_prices}
\end{table}

\begin{table}
 \begin{center}
  \pgfplotstabletypeset[
    string replace={0}{},
    fixed, fixed zerofill, precision=0, set thousands separator={},
    every head row/.style={before row=\toprule{\bf Certificate: times}&\multicolumn{3}{c|}{\bf Snowball}&\multicolumn{3}{c}{\bf Lock-in}\\\toprule,after row=\midrule},
    every last row/.style={after row=\bottomrule},
    every row no 3/.style={before row=\hline},
    every row no 6/.style={before row=\hline},
    every row no 9/.style={before row=\hline},
    columns={model,sn1t,sn2t,sn5t,li1t,li2t,li5t},
    columns/model/.style={string type,column type=l|,column name={\bf Model}},
    columns/sn1t/.style={column name={\bf 1y}},
    columns/sn2t/.style={column name={\bf 2y}},
    columns/sn5t/.style={column type=c|,column name={\bf 5y}},
    columns/li1t/.style={column name={\bf 1y}},
    columns/li2t/.style={column name={\bf 2y}},
    columns/li5t/.style={column name={\bf 5y}},
  ]{\certificatedat}
 \end{center}
 \caption[Computational time for snowball and lock-in certificates]{Algorithm computational time (in seconds) to price snowball and lock-in certificates with maturity of 1, 2 and 5 years (other characteristics described in the text). Comparison between different models (see text). LSMC models: polynomial bases, randomized neural networks (R-FFNN and R-RNN), signature based methods.}
 \label{tab:certificate_times}
\end{table}

Then, we investigate also the performance of finite-difference and Chebyshev methods in computing sensitivities. In Table~\ref{tab:gamma_snowball} we list the results. We analyze different algorithms, even if we have seen in Table~\ref{tab:certificate_prices} that prices are not in perfect agreement, to understand the stability of sensitivities. Polynomials basis function are taken up to third order; random neural networks have $120$ inner nodes; signature methods are truncated at third degree with lead-lag and time embedding. We can see that, for snowball certificates, both finite-difference and Chebyshev methods seem lead to similar results in term of stability. In Appendix~\ref{app:sensitivity} we extend the analysis.

\begin{table}
\begin{center}
  \pgfplotstabletypeset[
    string replace={0}{},
    fixed, fixed zerofill, precision=4, set thousands separator={},
    every head row/.style={after row=\midrule,before row=
      \toprule
      \multicolumn{9}{l}{\bf Snowball certificate: Gamma} \\
      \midrule
      & \multicolumn{4}{c|}{$\bm{S_0=85}$}
      & \multicolumn{4}{c}{$\bm{S_0=115}$} \smallskip\\
      \midrule
      & \multicolumn{2}{c|}{\small \bf Finite Difference}
      & \multicolumn{2}{c|}{\small \bf Chebyshev}
      & \multicolumn{2}{c|}{\small \bf Finite Difference}
      & \multicolumn{2}{c}{\small \bf Chebyshev} \smallskip\\
      },
    every last row/.style={after row=\midrule},
    columns={npath,ye085,fe085,yc085,fc085,ye115,fe115,yc115,fc115},
    columns/npath/.style={sci, precision=1, column name={\small \bf MC Path}},
    columns/ye085/.style={column name={\small \bf Polyn.}, 
    column type/.add={|}{}},
    columns/fe085/.style={column name={\small \bf FFNN}},
    columns/yc085/.style={column name={\small \bf Polyn.}, 
    column type/.add={|}{}},
    columns/fc085/.style={column name={\small \bf FFNN}},
    columns/ye115/.style={column name={\small \bf Polyn.}, 
    column type/.add={|}{}},
    columns/fe115/.style={column name={\small \bf FFNN}},
    columns/yc115/.style={column name={\small \bf Polyn.}, 
    column type/.add={|}{}},
    columns/fc115/.style={column name={\small \bf FFNN}},
  ]{\sensmcsnw}
  \pgfplotstabletypeset[
    string replace={0}{},
    fixed, fixed zerofill, precision=4, set thousands separator={},
    every head row/.style={after row=\midrule},
    every last row/.style={after row=\bottomrule},
    columns={npath,re085,se085,rc085,sc085,re115,se115,rc115,sc115},
    columns/npath/.style={sci, precision=1, column name={\small \bf MC Path}},
    columns/re085/.style={column name={\small \bf RNN}, 
    column type/.add={|}{}},
    columns/se085/.style={column name={\small \bf Sign.}},
    columns/rc085/.style={column name={\small \bf RNN}, 
    column type/.add={|}{}},
    columns/sc085/.style={column name={\small \bf Sign.}},
    columns/re115/.style={column name={\small \bf RNN}, 
    column type/.add={|}{}},
    columns/se115/.style={column name={\small \bf Sign.}},
    columns/rc115/.style={column name={\small \bf RNN}, 
    column type/.add={|}{}},
    columns/sc115/.style={column name={\small \bf Sign.}},
  ]{\sensmcsnw}
\end{center}
\caption[Gamma for a snowball certificate]{Gamma for a snowball certificate with maturity of two years (multiplied  by $10^4$). The first column is the total number of simulation paths used by each algorithm. Then, there are two groups each of four columns for different spot prices $S_0$. In each group the first two columns is the result of the finite-difference method ($\epsilon=1/4$) with different basis functions, the last two columns of the Chebyshev method ($\epsilon=3/4$).}
\label{tab:gamma_snowball}
\end{table}

\subsection{Lock-in payoff}
\label{subsec:lockin_num}

We consider a lock-in payoff with different maturities which pays quarterly a coupon if the underlying asset performance is above different coupon barriers. In particular we choose the following characteristics: (i) maturity of one year with coupon equal to $2.8\%$, coupon barrier at $100\%$ and capital barrier at $40\%$, (ii) maturity of two years with coupon equal to $2.4\%$, coupon barrier at $90\%$ and capital barrier at $30\%$, (iii) maturity of five years with coupon equal to $3\%$, coupon barrier at $90\%$ and capital barrier at $30\%$. The characteristics of the product are chosen so that the initial price is approximately at par, when evaluated with the LSMC method with polynomial basis on the risk set described at the beginning of Section~\ref{sec:callable}.

In the last three columns of Table~\ref{tab:certificate_prices} we report the prices of the lock-in certificates defined above by changing the basis functions used in the LSMC method to price the products. We can see that the prices are closely aligned, and within the Monte Carlo uncertainties, reported in appendix in Table~\ref{tab:certificate_errors}. In Table~\ref{tab:certificate_times} are listed the computational times. We can see that the best choices of basis functions are R-FFNN and R-RNN, the former being faster for shorter maturities but less accurate, followed by polynomials, although the last ones become unfeasible for many dates and larger degrees. Here, we consider more accurate a price which is lower than the others, since the dynamic program terminates the options to minimize the price.

Then, we investigate also the performance of finite-difference and Chebyshev methods in computing sensitivities. In Table~\ref{tab:gamma_lockin} we list the results. Polynomials basis function are taken up to third order; random neural networks have $120$ inner nodes; signature methods are truncated at third degree with lead-lag and time embedding. As in the previous case of snowball certificates, we obtain a similar behavior in term of stability both for finite-difference and Chebyshev algorithms. In Appendix~\ref{app:sensitivity} we extend the analysis.

\begin{table}
\begin{center}
  \pgfplotstabletypeset[
    string replace={0}{},
    fixed, fixed zerofill, precision=4, set thousands separator={},
    every head row/.style={after row=\midrule,before row=
      \toprule
      \multicolumn{9}{l}{\bf Lock-in certificate: Gamma} \\
      \midrule
      & \multicolumn{4}{c|}{$\bm{S_0=85}$}
      & \multicolumn{4}{c}{$\bm{S_0=115}$} \smallskip\\
      \midrule
      & \multicolumn{2}{c|}{\small \bf Finite Difference}
      & \multicolumn{2}{c|}{\small \bf Chebyshev}
      & \multicolumn{2}{c|}{\small \bf Finite Difference}
      & \multicolumn{2}{c}{\small \bf Chebyshev} \smallskip\\
      },
    every last row/.style={after row=\midrule},
    columns={npath,ye085,fe085,yc085,fc085,ye115,fe115,yc115,fc115},
    columns/npath/.style={sci, precision=1, column name={\small \bf MC Path}},
    columns/ye085/.style={column name={\small \bf Polyn.}, 
    column type/.add={|}{}},
    columns/fe085/.style={column name={\small \bf FFNN}},
    columns/yc085/.style={column name={\small \bf Polyn.}, 
    column type/.add={|}{}},
    columns/fc085/.style={column name={\small \bf FFNN}},
    columns/ye115/.style={column name={\small \bf Polyn.}, 
    column type/.add={|}{}},
    columns/fe115/.style={column name={\small \bf FFNN}},
    columns/yc115/.style={column name={\small \bf Polyn.}, 
    column type/.add={|}{}},
    columns/fc115/.style={column name={\small \bf FFNN}},
  ]{\sensmclck}
  \pgfplotstabletypeset[
    string replace={0}{},
    fixed, fixed zerofill, precision=4, set thousands separator={},
    every head row/.style={after row=\midrule},
    every last row/.style={after row=\bottomrule},
    columns={npath,re085,se085,rc085,sc085,re115,se115,rc115,sc115},
    columns/npath/.style={sci, precision=1, column name={\small \bf MC Path}},
    columns/re085/.style={column name={\small \bf RNN.}, 
    column type/.add={|}{}},
    columns/se085/.style={column name={\small \bf Sign.}},
    columns/rc085/.style={column name={\small \bf RNN}, 
    column type/.add={|}{}},
    columns/sc085/.style={column name={\small \bf Sign.}},
    columns/re115/.style={column name={\small \bf RNN}, 
    column type/.add={|}{}},
    columns/se115/.style={column name={\small \bf Sign.}},
    columns/rc115/.style={column name={\small \bf RNN}, 
    column type/.add={|}{}},
    columns/sc115/.style={column name={\small \bf Sign.}},
  ]{\sensmclck}
\end{center}
\caption[Gamma for a lock-in certificate]{Gamma for a lock-in certificate with maturity of two years (multiplied  by $10^4$). The first column is the total number of simulation paths used by each algorithm. Then, there are two groups each of four columns for different spot prices $S_0$. In each group the first two columns is the result of the finite-difference method ($\epsilon=1/4$) with different basis functions, the last two columns of the Chebyshev method ($\epsilon=3/4$).}
\label{tab:gamma_lockin}
\end{table}

\section{Conclusion and Further Developments}
\label{sec:conclusion}

In the present paper, we studied state of the art algorithms, that are now classified under the name of machine learning, to price Asian, look-back products, and callable certificates with early-termination features. More precisely, we adapted the approach of~\textcite{herrera2021optimal} to the case of path-dependent payoffs with early-termination features, and we compare the results with alternative original formulations based on signature methods. All the experiments are run under a Black-Scholes dynamics for a single stock price.

We observed that, at least for the type of payoffs and the dynamics we considered, these algorithms have a performance, in terms of accuracy and computational efficiency, often comparable to that of more traditional algorithms, such as the ones based on polynomial basis functions. 
In particular, we observed that a careful selection of the risk factors in traditional approaches may lead to relevant improvement in computational time by saving the accuracy of the results, at least when pricing Asian and look-back products with early-termination features. 
On the other hand, when pricing callable certificates, the best choices of basis functions are random networks (R-FFNN and R-RNN); in this case, polynomials become unfeasible for many dates and larger degrees. 
In this context, signatures do not emerge as strictly necessary algorithms. 
Although prices obtained from truncated signatures are among the highest for look-back options and certificates, these are matched by R-FFNNs which, moreover, have much lower computational times. 
Further, more investigations would be needed to address the computational issues (inversion of the regression basis) which turns out to be (almost) singular\footnote{This phenomenon is already known in the literature and is due to the fact that some signature coefficients of embedded paths may be equal (see, for example, Section 2.1.5 in \textcite{primer_signature_ML} - version 1).} with regularization techniques (e.g.~Lasso or Ridge regularization). 
Such problems can be avoided using randomized signatures, as alluded to in Section \ref{sec:rand_signature}, for which we can obtain similar, though not equal, results. 
Standard signatures show their abilities as feature extractors particularly well for high truncations, and that is where advantages compared to randomized signatures become more apparent, but the latter have the advantage of being able to offer similar expressiveness at lower computational costs.\\

As a second contribution, we analyzed algorithms to compute sensitivities; in particular, we found that Chebyshev interpolation techniques are an effective choice for Delta and Gamma calculations. 
The strength of the method relies in decoupling the selection of the interpolation interval from the computation of the derivatives, so that a larger size of the interval, required to stabilize the sensitivity computation, does not introduce biases. 
To the best of our knowledge, it is the first time that first and second order Greeks are computed for Amerasian derivatives and callable certificates.\\

As further developments, we plan to investigate the performance of machine learning based methods for more complicated, e.g., non-Markov dynamics, and in higher dimensional settings.

\section*{Acknowledgements}

All the authors warmly thank Antonino Zanette (University of Udine) for having provided the MATLAB code for the implementation of the benchmark methodologies used for Asian payoffs. They thank Qi Feng (Florida State University) for having provided insights on the usefulness or not of the lead-lag transformation. They thank also Francesca Sivero for fruitful discussions on design patterns and Giuseppe Di Poto for American options calculations with binomial trees.

\section*{Disclaimer}

The authors report no potential competing interests. The opinions expressed in this document are solely those of the authors and do not represent in any way those of their present and past employers.

\printbibliography

\newpage

\appendix

\section{The GPR-GHQ algorithm and the binomial Markov chain method}
\label{app:Zanette}

We discuss both the GPR-GHQ algorithm and the binomial Markov chain method presented in~\textcite{Goudenege2022}. We start with the description of the GPR-GHQ algorithm. We remind that GPR stands for Gaussian Process Regression and GHQ for Gauss-Hermite quadrature. It is a backward induction algorithm that employs the GHQ to compute the continuation value of the option only for some particular path of the underlying and the GPR to extrapolate the whole continuation value from those observations.

\subsection{Details on the GPR-GHQ algorithm}
\label{app:gpr-ghq}

Precisely, let $N$ be the number of time steps and $T_{i}$ as in Subsection~\ref{sec:amerasian}. In addition, let $(\mathbf{A}_{i})_{M\leq i\leq N}$ and $(\mathbf{B}_{i})_{M\leq i\leq N}$ be the following two processes:
\begin{equation*}
    \mathbf{A}_{i}=(\mathbf{A}_{i,1},\ldots,\mathbf{A}_{i,d_{i}^A})^T = (A_{i-M+1}^{i}, A_{i-M+2}^{i},\ldots,A_{\min\{i-1,N-M+1\}}^{i}, A_{i}^{i})^T
\end{equation*}
and
\begin{equation*}
   \mathbf{B}_{i}=(\mathbf{B}_{i,1},\ldots,\mathbf{B}_{i,d_{i}^B})^T = (A_{i-M+2}^{i}, A_{i-M+3}^{i},\ldots,A_{\min\{i-1,N-M+1\}}^{i}, A_{i}^{i})^T.
\end{equation*}
In the previous equations, $A_{i_1}^{i_2}$, with $i_1$ and $i_2$ two natural numbers, denotes a quantity closely related to the quantity introduced in Equation~\eqref{eq:movingaverage}
\begin{equation*}
    A_{i_1}^{i_2}:=\frac{1}{i_2-i_1+1}\sum_{j=i_1}^{i_2} S_{T_j}.
\end{equation*}

In particular, $\mathbf{B}_{i}$ can be obtained from $\mathbf{A}_{i}$ by dropping the first component $\mathbf{A}_{i,1}$. Let $\mathcal{V}_{i}$ be the option value at time $T_{i}$, which is determined by the process of partial averages $\mathbf{A}_{i}$ in the following way  
\begin{equation*}
    \mathcal{V}_{i}(\mathbf{A}_{i}) = \max(\Psi_{i}^{\mathbf{A}}(\mathbf{A}_{i}), C_{i}(\mathbf{B}_{i})),
\end{equation*}
with $C_{i}$ the continuation value function of the moving average option at time $T_{i}$, where we explicitly write the dependence upon the process $\mathbf{B}_{i}$. The latter is given by the following relation:
\begin{equation*}
    C_{i}(\mathbf{B}_{i}):=\mathds{E}_{T_i,\mathbf{B}_{i}}\left[e^{-r\Delta t}\mathcal{V}_{i+1}(\mathbf{A}_{i+1})\right],
\end{equation*}
where $\mathds{E}_{T_i,\mathbf{B}_{i}}$ represents the expectation at time $T_i$ given that $\mathbf{B}_{i}$ is the value of the process $\mathbf{B}$ at time $T_i$. Moreover, $\Psi_{T_i}^{\mathbf{A}}$ denotes the payoff as a function of the process $\mathbf{A}_{i}$
\begin{equation*}
    \Psi_{T_i}^{\mathbf{A}}(\mathbf{A}_{i}):=\max(0, A_{i}^{i}-A_{i-M+1}^{i}).
\end{equation*}

Then, the dynamic programming problem of the function $\mathcal{C}$ is given by (cfr.~Equation (4.10) of~\textcite{Goudenege2022}):
\begin{equation*}
    \begin{cases}
        C_N(\mathbf{B}_{N})=0,\\
        C_{N-1}(\mathbf{B}_{N-1})=\text{Call}\left(t_{N-1}, t_{N}, A_{N-1}^{N-1}, \frac{M A_{i-M+1}^{i}-A_{i}^{i}}{M-1}\right),\\
        C_{i}(\mathbf{B}_{i})=\mathds{E}_{T_{i},\mathbf{B}_{i}}\left[e^{-r\Delta t}\max\left(\Psi_{i+1}^A(\mathbf{A}_{i+1}),C_{i+1}(\mathbf{B}_{i+1})\right)\right],
    \end{cases}
\end{equation*}
where $N-2 \leq i \leq M$ and $\text{Call}(t_0, T, S_0, K)$ is the price of a European call option on $S$ with inception time $t_0$, maturity $T$, spot value $S_0$, and strike $K$. At this point, \textcite{Goudenege2022} exploit GPR to learn $C_{i}$ from a few observed values. Toward this aim, they consider a set $X^{i}$ of $P$ points whose elements are the simulated values for $\mathbf{B}_{i}$. More precisely:
\begin{equation*}
    X^{i}=\{\mathbf{x}^{i,p}=(x_1^{i,p},\ldots,x_{d_{i}^B}^{i,p}),\,p=1,\ldots,P\}\subseteq\mathds{R}^{d_{i}}.
\end{equation*}

The GPR-GHQ method determines $C_{i}(\mathbf{x}^{i,p})$ for each $\mathbf{x}^{i,p}\in X^{i}$ through $Q$-points GHQ by exploiting the following relation
\begin{equation*}
    C_{i}^{GHQ}(\mathbf{x}^{i,p})=e^{-r\Delta t}\sum_{q=1}^{Q}w_q\max\left(\Psi_{i+1}^{A}(\tilde{x}_0^{i,p,q},\tilde{\mathbf{x}}^{i,p,q}),\mathcal{C}_{i+1}(\tilde{\mathbf{x}}^{i,p,q})\right),
\end{equation*}
where $\tilde{x}_{d_{i+1}}^{i,p,q}=S^{i,p,q}$, $\tilde{x}_{i}^{n,p,q}=\frac{(M-i-1)x_{i+1}^{n,q}+S^{n,p,q}}{M-i}$, and $\tilde{x}_{0}^{n,p,q}=\frac{(M-1)x_{1}^{n,q}+S^{n,p,q}}{M}$. In particular, the vector $(\tilde{x}_{0}^{i,p,q}, \tilde{\mathbf{x}}^{i,p,q})$ is a possible outcome for $\mathbf{A}_{i+1}|\mathbf{B}_{i} = \mathbf{x}^{i,p}$. The above equation, can be evaluated only if the quantities $C_{i+1}(\tilde{\mathbf{x}}^{i,p,q})$ are known for all the future points $\tilde{\mathbf{x}}^{i,p,q}$. In order to do so, \textcite{Goudenege2022} employ the GPR method by leading to the subsequent equation:
\begin{equation}
    C_{i}^{GPR-GHQ}(\mathbf{x}^{i,p})=e^{-r\Delta t}\sum_{q=1}^{Q}w_q\max\left(\Psi_{i+1}^A(\tilde{x}_{0}^{i,p,q}, \tilde{\mathbf{x}}^{i,p,q}),C_{i+1}^{GPR}(\tilde{\mathbf{x}}^{n,p,q})\right),\,\,p\in\{1,\ldots,P\}.
\end{equation}

Once the function $C_{T_i}^{GPR-GHQ}$ has been estimated, the option price $\mathcal{V}_0$ at inception can be computed by discounting the expected option value at time $T_M$, which is the first time step the option can be exercised; \textcite{Goudenege2022} employ a Monte Carlo approach with antithetic variables to estimate the expected value involved in the latter step.

\subsection{Details on the binomial Markov chain method}

Now, we describe the binomial chain. \textcite{Goudenege2022} consider a Markov chain defined on a Cox-Ross-Rubinstein (CRR, henceforth) binomial tree. The set of all possible states at time $T_{i}$ for $M \leq i$ is given by
\begin{equation*}
    \{(\mathbf{s}_p, S_{T_i,k}),\,p=1,\ldots,2^{M-1},\,k=0,\ldots,i\},
\end{equation*}
where $\mathbf{s}_p=(\mathbf{s}_{p,1},\ldots, \mathbf{s}_{p,M-1})^{T}$ and $S_{T_i,k}=S_0 e^{(2k-i)\sigma\sqrt{\Delta t}}$. In particular, the payoff for a state $(\mathbf{s}_p, S_{T_i,k})$ becomes:
\begin{equation}
    \Psi^{CRR}(\mathbf{s}_p, S_{T_i,k})=\max\left(S_{T_i,k}-\frac{1}{M}\sum_{j=1}^{M}S_{j,k}\exp\left(-\sigma\sqrt{\Delta t}\sum_{\ell=M-j}^{M-1}(2\mathbf{s}_{p,\ell}-1\right)\right).
\end{equation}

In particular, if the process state at time $T_i$ is $(\mathbf{s}_p, S_{T_i,k})$, then the possible next states are denoted with $(\mathbf{s}_p^{\text{up}}, S_{T_i,k}^{\text{up}})$ and $(\mathbf{s}_p^{\text{up}}, S_{T_i,k}^{\text{dw}})$; option evaluation is performed by moving backward along the tree. By exploiting the positive homogeneity of the continuation value, denoted by $C_{T_i}^{CRR}$, \textcite{Goudenege2022} arrive at the following recursive formula:
\begin{equation}
  \begin{cases}
    C_N^{BC}(\mathbf{s}_p)=0,\medskip\\
    \!\begin{aligned}
      C_N^{BC}(\mathbf{s}_p) =& \;e^{-r\Delta t}\left[p_{\text{up}}e^{\sigma\sqrt{\Delta t}}\max(\Psi^{CRR}(\mathbf{s}_p^{\text{up}},1),\mathcal{C}^{BC}_{i+1}(\mathbf{s}_p^{\text{up}}))\right. \\
      & \left. + \,p_{\text{dw}}e^{-\sigma\sqrt{\Delta t}}\max(\Psi^{CRR}(\mathbf{s}_p^{\text{dw}},1),C^{BC}_{i+1}(\mathbf{s}_p^{\text{dw}}))\right].
    \end{aligned}
  \end{cases}
\end{equation}

Finally, once the continuation value is available at time step $M$ the option value at inception is obtained by averaging the option value at the various states of the binomial chain at time $T_M$. The total number of possible states is $O(N 2^M)$ and the computational cost is $O(N 2^{M+1})$. 

\section{Signatures methods}
\label{app:signature}

This section collects some basic concepts and definitions on signatures. In particular, Subsection~\ref{app::standardsignature} provides background material on signatures, with an adaptation of the signature transform to the space of streams of data, and Subsection~\ref{app:numerical_aspects} discusses some numerical aspects of signatures. 

\subsection{Background material on signatures}
\label{app::standardsignature}

We start with the definition of the signature of a path. Let $X:J\rightarrow\mathds{R}^d$ be a $d$-dimensional path where $J$ is a compact interval. A path $X$ is of finite $p$-variation for certain $p \geq 1$ if the $p$-variation of $X$, defined by
\begin{equation*}
  \|X\|_{p,J}=\left(\sup_{D_J\subset J}\sum_{i}\|X_{T_i}-X_{T_{i-1}}\|^p\right)^{1/p}
\end{equation*}
is finite. In the previous equation, the supremum is taken over all possible finite partitions $D_J=\{T_i\}_{i}$ of the interval $J$. 
We let $\mathcal{V}^{p}(J, E)$, with $E \subseteq \mathds{R}^{d}$, be the set of any continuous path $X:J\rightarrow E$ of finite $p$-variation.

\begin{definition}[The signature of a path]
Let $J$ be a compact interval and $X \in \mathcal{V}^{p}(J, E)$ such that the integration in Equation~\eqref{eq::truncatedsignature} makes sense. The signature $S(X)$ of $X$ over the time interval $J$ is defined as follows
\begin{equation*}
   S(X)=\mathbf{X}_{J} = (1, X^1, \ldots, X^n, \ldots),
\end{equation*}
where for each integer $n\geq 1$,
\begin{equation}\label{eq::truncatedsignature}
   X^n = \left({\int\ldots\int}_{\substack{u_1<\ldots<u_n \\ u_1,\ldots,u_n\in J}} dX_{u_1}\otimes\ldots\otimes dX_{u_n}\right)\in E^{\otimes n},
\end{equation}
with $\otimes$ denoting the tensor product.
\end{definition}

Let us now take $E=\mathds{R}^d$ for simplicity. We notice that the $n=0$ term is $1 \in (\mathds{R}^{d})^{\otimes 0}=\mathds{R}$, the first term belongs to $\mathds{R}^d$, the second term belongs to $\mathds{R}^{d} \otimes \mathds{R}$ (that is, the space of matrices of size $d\times d$), and, in general, the $k^{th}$ term belongs to $(\mathds{R}^d)^{\otimes k}=\mathds{R}^d \otimes \ldots \otimes \mathds{R}^{d}$, $k$ times (that is, the space of tensors of shape $(d,\ldots,d)$, $k$ times. In particular, it turns out that the signature of $X$ is an element of the tensor algebra space $T((\mathds{R}^d))$, which is also the ``free'' algebra on $\mathds{R}^d$, defined as 
\begin{equation}\label{eq:tensor_algebra}
  T((\mathds{R}^d)):=\left\{(a_0, a_1, \ldots, a_n, \ldots) \,|\, \forall n \geq 0,\,a_n \in (\mathds{R}^d)^{\otimes n}\right\}.
\end{equation}
We notice that \eqref{eq:tensor_algebra} can also be written as 
\begin{equation*}
  T((\mathds{R}^d)):=\{ a \,|\, a = \sum_{k=0}^{+\infty} \sum_{i_1, \dots, i_k =1}^d a_{i_1 \dots i_k} e_{i_1} \cdots e_{i_k} \},
\end{equation*}
where $e_i = (0, \ldots, 0, 1, 0, \ldots, 0)$ with 1 in the $i$\textsuperscript{th} position, are elements of the basis of $\mathds{R}^d$.

\subsection{Numerical aspects of signatures}
\label{app:numerical_aspects}

The truncated signature of $X$ of order $n$ is denoted by $S^{n}(X)$, i.e., $S^{n}(X)=(1, X^{1}, \ldots, X^{n})$, for every integer $n \geq 1$. The truncated signatures of $X$ of order $n$ lies in $T^{n}((\mathds{R}^d)):= \bigotimes_{i=0}^{n} E^{\otimes i}$. In particular, the signature truncated at order $n$ is a vector of dimension 
\begin{equation}
\label{eq: sig_dim}
  s_d(n) = \sum_{k=0}^{n} d^{k} = \frac{d^{n+1}-1}{d-1},
\end{equation}
if $d \geq 2$, $s_d(n) = n+1$ if $d = 1$. The Python software \texttt{iisignature} that we used for our calculations ignores the constant number 1 (first element of any signature) and, thus, the dimension reduces to $s_d(n)-1$.

Given a path $X \in \mathcal{V}^{p}(J,E)$, with $p\geq 1$, we know that the signature of the path uniquely defines (up to tree-like equivalence) the path itself. This explains why it can be so convenient to ``summarize'' a path in terms of its signatures coefficients: essentially, no information is lost. This result for paths of bounded variation is due to~\textcite{hambly2010uniqueness} and has been extended by~\textcite{Boedihardjo2016} to geometric rough paths, to which we refer for the definition of tree-like equivalence as well.
\begin{theorem}[Uniqueness of Signature]
  Let $X \in \mathcal{V}^{p}([0,T], \mathds{R}^{d})$, $p\geq 1$. Then $S(X)$ determines $X$ up to the tree-like equivalence.
\end{theorem}

To avoid having degenerate paths (in a tree-like sense), it has become customary to define the so-called time-augmented path version of a path by $\widehat{X}_t = (t, X_t)$, which is a path in $\mathds{R} \times E$. Indeed, the following proposition holds true.
\begin{proposition}[Uniqueness of signatures]
  Let $X \in \mathcal{V}^p(J,E)$ and $\widehat{X}$ its time-augmented version. Then $S(\widehat{X})$ uniquely determines $X$ up to translation.
\end{proposition}

A relevant property of signatures is given by the following Proposition, which states that the terms of the signature decay in size in a factorial way; see Lemma~2.1.1 in~\textcite{lyons1998differential}.

\begin{proposition}[Decay of signature terms]
\label{prop::factorial}
  Let $X \in \mathcal{V}^{p}(J, E)$. Then 
  \begin{equation*}
    \| X^n \| \leq \frac{C(X)^n}{n!},
  \end{equation*}
  where $X^n$ is defined in Equation \eqref{eq::truncatedsignature}, $C(X)$ is a constant depending on $X$ and $\|\cdot\|$ is any tensor norm on $(\mathds{R}^d)^{\otimes n}$.
\end{proposition}

Furthermore, we have to cite another relevant property of signatures, namely that the signature is invariant to time reparametrization; see~\textcite{lyons1998differential}.

\begin{proposition}[Invariance to time reparametrization]
\label{prop::composition}
  Let $X \in \mathcal{V}^p(J, E)$. Let $\psi: E \rightarrow E$ be continuously differentiable, increasing, and surjective. Then $S(X)=S(X \circ \psi)$, where $ ``\circ"$ denotes the composition.
\end{proposition}

In practice, we interpret a stream of data as a discretization of a path. The space of streams of data is defined as:
\begin{equation*}
  \text{Str}(E) = \{\boldsymbol{x}=(x_1, x_2, \ldots, x_N)\,:\,x_i \in \mathds{R}^d,\,N\in\mathds{N}\}.
\end{equation*}
Given $\boldsymbol{x} \in \text{Str}(E)$, the integer $N$ is called the length of $\boldsymbol{x}$. Furthermore, for $a, b \in \mathds{R}$ such that $a<b$, fix
\begin{equation*}
  a = u_1 < u_2 < \ldots < u_{N-1} < u_N = b.
\end{equation*}

Let $J = [a,b]$ and $X = (X^1,\ldots,X^d) : J \rightarrow E$ be in $\mathcal{V}^p(J, E)$ such that $X_{u_i} = x_i$ for all $i$, and interpolated in continuously differentiable, increasing, and surjective way on the intervals in between. Then $X$ is called interpolation of $\boldsymbol{x}$. We have the following definition.

\begin{definition}[Signature of the interpolating path]
Let $\boldsymbol{x} \in \text{Str}(E)$. Let $X$ be an interpolation of $\boldsymbol{x}$.
Then the signature of $\boldsymbol{x}$ is defined as $S(\boldsymbol{x}) = S(X)$ and similarly the truncated signature. 
\end{definition}

\noindent In particular, thanks to Proposition~\ref{prop::composition} the previous definition is well defined and independent of the interpolation choice.

Finally, one of the most important properties of signatures is being universal approximators, that is continuous functions on paths can be approximated on compact spaces by linear functions on the signature of the path; see, for instance, \textcite{Kirali2019}.
\begin{theorem}[Universal approximators]\label{thm: sig-universal-approx}
    Let $K$ be a compact set in $\mathcal{V}^{p}([0,T], \mathds{R}^{d})$ and $f \in \mathcal{C}(K, \mathds{R}^d)$ be an $\mathds{R}^d$-valued continuous function on such compact.
    Then, for any $\epsilon>0$, there exists a linear function $l$ such that $\sup_{X\in K }\| f(X) - \langle l, S(\widehat{X}) \rangle \| < \epsilon$. 
\end{theorem}

\section{Snowball and lock-in payoffs}
\label{app:callable}

A certificate is a financial instrument issued by a financial intermediary that allows to take a position on one or many underlying assets. Here, we focus on certificates with maturity date $T$ written on a single traded assets, whose price process is $S_t$, and paying coupons $\gamma_i$ on payment dates $T_i$ with $1 \le i \le N$, $T_1>0$ and $T_N:=T$. On the certificate maturity date the whole principal amount, or a fraction of it, is redeemed according to the performance of the underlying asset. The principal-amount redemption is defined as
\begin{equation}
  \phi(s,H) := \ind{ P(s) > H } + \ind{ P(s) < H } P(s) ,
\end{equation}%
where $P$ is the basket worst performance and it is given by
\begin{equation}
  P(s) := \frac{s}{s_0}
\end{equation}%
with respect to a level $s_0$ usually read from the market before issuing the certificate.

In case the contract is terminated before the maturity the whole principal amount is redeemed. This case may happen if the contract contains early-termination features, as discussed in the next section. For later convenience, we define the principal-amount redemption on maturity date or on early-termination date as following
\begin{equation}
 \Psi^{\rm C}_{T_i}(N;H) := 1 - \ind{i=N} \left( 1 - \phi(S_{T_N},H) \right).
\end{equation}

The coupon amounts paid by the certificate can be fixed at inception or they can depend on the asset performances. We wish to investigate the latter case, and, in particular, the case of coupons with ``snowball'' and ``lock-in'' features.

We start by discussing the ``snowball'' version. At each payment date the certificate pays a coupon only if the basket worst performance is above a barrier level $K$. When the payment occurs, the quantity of cash paid is equal to the sum of the cash flows not yet paid. We can write this feature in the following way. First, we introduce the set ${\cal J}(i)$ of indices less than $i$ such that a coupon can be paid on the corresponding payment date without taking into account if the product is early terminated. In formulae we have
\begin{equation}
  {\cal J}(i) := \{ j \in \{1,\ldots,i-1\} : P(S_{t_j})>K \}.
\end{equation}%
We can now define the coupon as
\begin{equation}
  \gamma_i := \ind{P(S_{t_i})>K} \,\sum_{j=1+\sup{\cal J}(i)}^i c_j 
\end{equation}%
with the convention that the supremum of ${\cal J}(i)$ is $0$ when the set is empty. In the previous equation $\{c_1,\ldots,c_N\}$ is a pre-determined set of non-negative cash flows.

Then, we discuss the ``lock-in'' version. At each payment date the certificate pays a coupon only if the basket worst performance at such date, or at any of the previous payment dates, is above a barrier level $K$. We can define the coupon as
\begin{equation}
  \gamma_i := \max_{j\in[1,i]} \ind{P(S_{t_j})>K} \,C_{T_i} .
\end{equation}%

We can evaluate certificates with the possibility of an early exercise by the issuer. Again, we consider that the certificate can be exercised on any date in the time grid $\{T_1, \ldots, T_N\}$ previously defined. We setup the pricing problem by starting from the last exercise date $T_N$, and we introduce the ${\cal F}_{T_N}$-measurable value function $V^{\rm C}_N$, defined as
\begin{equation}
  V^{\rm C}_{T_N} := \Psi^{\rm C}_{T_N} .
\end{equation}%
On the previous dates $T_i$, with $0<i<N$ the option buyer may decide to exercise the option, by choosing the minimum between the immediate redemption of the principal amount and the continuation value of the contract. Thus, we can write
\begin{equation}
\label{eq:cert_dynprog}
 V^{\rm C}_{T_i} := \gamma_i + \min \left\{ \Psi^{\rm C}_{T_i}, B_{T_i} \,\ExF{T_i}{ \frac{V^{\rm C}_{T_{i+1}}}{B_{T_{i+1}}} } \right\} ,
\end{equation}%
The price of the certificate can be calculated by applying the above recursion up to $T_1$ and by calculating the certificate price as given by
\begin{equation}
  V^{\rm C}_0 := \ExO{\frac{V^{\rm C}_{T_1}}{B_{T_1}}} .
\end{equation}%

\section{Technical insights on sensitivity computations}
\label{app:sensitivity}

Here, we describe in more details the problem of dealing with discontinuous sensitivities. In particular, we expand the discussion on the Chebyshev method presented in Section~\ref{sec:chebyshev} in the main text of the paper. Then, we conclude this section by presenting more numerical evidences supporting our choice for the Chebyshev method.

\subsection{Discontinuous sensitivities in the Chebyshev method}
\label{app:sensitivity-jump}

In the case of American put options the buyer may exercise the option on contract inception. As a consequence the function $S \mapsto V(S)$ has a discontinuous second derivative. In our specific settings of positive interest rates and zero dividend yield (see Section~\ref{sec:dynamics}), we know that the immediate exercise occurs when the spot price is less than a critical value $S^\star$, which depends on market and contract data. Moreover, when it occurs, the option price is equal to its intrinsic value, a linear function of the underlying spot price, leading to a Gamma equal to zero.

We should perform the numerical calculation only when the spot price is greater than $S^\star$. In the case of the Chebyshev method we can achieve this by redefining the grid of spot prices, previously defined in \eqref{eq:chebpoint}, as
\begin{equation}
S^\ell_\epsilon := \max\left\{ S , \frac{S^\star}{1-\epsilon} \right\} \left(1 + \epsilon \cos\left(\frac{\ell\pi}{N_c-1}\right)\right) ,
\label{eq:chebpoint_star}
\end{equation}%
where $S>S^\star$ is the market spot price.

The above derivation requires the knowledge of $S^\star$. However, if the value of $S^\star$ is not known in advance, we can implement a simple heuristic.
\begin{enumerate}
 \item We choose $N_c$ to be odd, namely $N_c:=2K_c+1$ for a positive integer $K_c$, and we set $S^\star:=S(1-\epsilon)$ in equation~\eqref{eq:chebpoint_star}, so that $S^{K_c}_\epsilon=S$.
 \item We calculate the corresponding Monte Carlo prices and we check if on $S^{K_c}_\epsilon$ an immediate exercise occurs. If it is the case we return zero for the Gamma, otherwise we go on.
 \item We check if an immediate exercise occurs for some $S^\ell_\epsilon$ with $\ell<K_c$. If this is the case we discard such points from the interpolator training set.  
 \item We train the interpolator on the remaining points, and we calculate the sensitivity.
\end{enumerate}

\subsection{Numerical investigations on American put sensitivities}
\label{app:sensitivity-put}

We start by presenting the results in Figure~\ref{fig:american_put_sensitivity_fd} Delta and Gamma in the case of the finite difference method for American put options at different spot prices. We can see that increasing $\epsilon$ the noise is reduced, but a bias in the sensitivity arises, since only in the limit of vanishing $\epsilon$ we recover the true derivative values.

\begin{figure}
\ifx\RemoveFigures\undefined
  \begin{center}
  \scalebox{0.95}{%
    \begin{tikzpicture}
      \begin{axis}[set layers, mark layer=axis background,
                  xlabel=Spot Price,
                  ylabel=Delta,
                  ylabel style={overlay},
                  xmin=0.55, xmax=1.45,
                  xtick={0.6,0.8,1,1.2,1.4},
                  xticklabels={60,80,100,120,140},
                  ymin=-1.05, ymax=0.05,
                  grid=major,
                  legend style={legend pos=north west},
                  axis background/.style={fill=gray!10}]
      \addplot [color=mydarkblue!39,thick,smooth] table [y=e32d,x=money] from \senscmp;
      \addplot [color=mydarkblue!69,thick,smooth] table [y=e16d,x=money] from \senscmp;
      \addplot [color=mydarkblue!99,thick,smooth] table [y=e8d,x=money] from \senscmp;
      % \addplot [color=mydarkblue!99,thick,smooth] table [y=e4d,x=money] from \senscmp;
      % \addplot [color=mydarkblue!59,thick,smooth] table [y=etd,x=money] from \senscmp;
      % \addplot [color=mydarkblue!79,thick,smooth] table [y=emd,x=money] from \senscmp;
      % \addplot [color=mydarkblue!99,thick,smooth] table [y=ecd,x=money] from \senscmp;
      \addplot [color=mydarkorange,thick,only marks,mark options={scale=.5}] table [y=btd,x=money] from \senscmp;
      % \legend{{\small $\epsilon=\sfrac{1}{16}$},{\small $\epsilon=1/8$},{\small $\epsilon=1/4$},{\small BT}}
      \legend{{\small $\epsilon=\sfrac{1}{32}$},{\small $\epsilon=\sfrac{1}{16}$},{\small $\epsilon=1/8$},{\small BT}}
      \end{axis}
    \end{tikzpicture}
    ~~~
    \begin{tikzpicture}
      \begin{axis}[set layers, mark layer=axis background,
                  xlabel=Spot Price,
                  ylabel=Gamma,
                  ylabel style={overlay},
                  xmin=0.55, xmax=1.45,
                  xtick={0.6,0.8,1,1.2,1.4},
                  xticklabels={60,80,100,120,140},
                  ymin=-0.005, ymax=0.035,
                  grid=major,
                  legend style={legend pos=north west},
                  axis background/.style={fill=gray!10}]
      \addplot [color=mydarkblue!39,thick,smooth] table [y=e32g,x=money] from \senscmp;
      \addplot [color=mydarkblue!69,thick,smooth] table [y=e16g,x=money] from \senscmp;
      \addplot [color=mydarkblue!99,thick,smooth] table [y=e8g,x=money] from \senscmp;
      % \addplot [color=mydarkblue!99,thick,smooth] table [y=e4g,x=money] from \senscmp;
      % \addplot [color=mydarkblue!59,thick,smooth] table [y=etg,x=money] from \senscmp;
      % \addplot [color=mydarkblue!79,thick,smooth] table [y=emg,x=money] from \senscmp;
      % \addplot [color=mydarkblue!99,thick,smooth] table [y=ecg,x=money] from \senscmp;
      \addplot [color=mydarkorange,thick,only marks,mark options={scale=.5}] table [y=btg,x=money] from \senscmp;
      \end{axis}
    \end{tikzpicture}}%
  \end{center}
\fi
\caption[Delta and Gamma risk matrices for American put options -- finite differences]{Delta and Gamma risk matrices for American put options with maturity $T=0.2$, strike $K=100$ and time steps $N=50$. Orange dots are calculated with a standard binomial tree implementation, while blue lines with the finite difference method.}
\label{fig:american_put_sensitivity_fd}
\end{figure}

Then, we continue with the regression method and we show in Figure~\ref{fig:american_put_sensitivity_rn} the calculation of Gamma for American put options at different spot prices by varying both $\epsilon$ and $B$. We can see that a delicate fine tuning is required to properly select these variables to reduce noise without introducing a bias. In particular, we notice that higher terms in the Taylor's approximation are required to correctly learn the dependency of the derivative price with respect to the spot price, but they increase the sensitivity noise. In order to reduce the noise we can increase the variance of the spot-price density, but in this way we introduce a bias. Moreover, we can see that the discontinuity of Gamma cannot be reproduced with this method, since the approximating form for the derivative price is continuous in the spot price.

\begin{figure}
\ifx\RemoveFigures\undefined
  \begin{center}
  \scalebox{0.95}{%
    \begin{tikzpicture}
      \begin{axis}[set layers, mark layer=axis background,
                  xlabel=Spot Price,
                  ylabel=Gamma,
                  ylabel style={overlay},
                  xmin=0.55, xmax=1.45,
                  xtick={0.6,0.8,1,1.2,1.4},
                  xticklabels={60,80,100,120,140},
                  ymin=-0.005, ymax=0.035,
                  grid=major,
                  legend style={legend pos=north east},
                  axis background/.style={fill=gray!10}]
      % \addplot [color=mydarkblue!39,thick,smooth] table [y=r32g,x=money] from \senscmp;
      \addplot [color=mydarkblue!39,thick,smooth] table [y=r16g,x=money] from \senscmp;
      \addplot [color=mydarkblue!69,thick,smooth] table [y=r8g,x=money] from \senscmp;
      \addplot [color=mydarkblue!99,thick,smooth] table [y=r4g,x=money] from \senscmp;
      % \addplot [color=mydarkblue!59,thick,smooth] table [y=rtg,x=money] from \senscmp;
      % \addplot [color=mydarkblue!79,thick,smooth] table [y=rmg,x=money] from \senscmp;
      % \addplot [color=mydarkblue!99,thick,smooth] table [y=rcg,x=money] from \senscmp;
      \addplot [color=mydarkorange,thick,only marks,mark options={scale=.5}] table [y=btg,x=money] from \senscmp;
      \legend{{\small $\epsilon=\sfrac{1}{16}$},{\small $\epsilon=1/8$},{\small $\epsilon=1/4$},{\small BT}}
      \end{axis}
    \end{tikzpicture}
    ~~~
    \begin{tikzpicture}
      \begin{axis}[set layers, mark layer=axis background,
                  xlabel=Spot Price,
                  ylabel=Gamma,
                  ylabel style={overlay},
                  xmin=0.55, xmax=1.45,
                  xtick={0.6,0.8,1,1.2,1.4},
                  xticklabels={60,80,100,120,140},
                  ymin=-0.005, ymax=0.035,
                  grid=major,
                  legend style={legend pos=north east},
                  axis background/.style={fill=gray!10}]
      % \addplot [color=mydarkblue!39,thick,smooth] table [y=r129g,x=money] from \senscmp;
      \addplot [color=mydarkblue!39,thick,smooth] table [y=r39g,x=money] from \senscmp;
      \addplot [color=mydarkblue!69,thick,smooth] table [y=r69g,x=money] from \senscmp;
      \addplot [color=mydarkblue!99,thick,smooth] table [y=r99g,x=money] from \senscmp;
      % \addplot [color=mydarkblue!39,thick,smooth] table [y=r12g,x=money] from \senscmp;
      % \addplot [color=mydarkblue!59,thick,smooth] table [y=rmg,x=money] from \senscmp;
      % \addplot [color=mydarkblue!79,thick,smooth] table [y=r6g,x=money] from \senscmp;
      % \addplot [color=mydarkblue!99,thick,smooth] table [y=r3g,x=money] from \senscmp;
      \addplot [color=mydarkorange,thick,only marks,mark options={scale=.5}] table [y=btg,x=money] from \senscmp;
      \legend{{\small $B=3$},{\small $B=6$},{\small $B=9$},{\small BT}}
      \end{axis}
    \end{tikzpicture}}%
  \end{center}
\fi
\caption[Gamma risk matrices for American put options -- regression]{Gamma risk matrices for American put options with maturity $T=0.2$, strike $K=100$ and time steps $N=50$. Orange dots are calculated with a standard binomial tree implementation, while blue lines with the regression method. Left panel: different choices of $\epsilon$ for $B=9$. Right panel: different choices of $B$ for $\epsilon=1/8$.}
\label{fig:american_put_sensitivity_rn}
\end{figure}

Finally, we analyze the Chebyshev method and we present in Figure~\ref{fig:american_put_sensitivity_cheb} the calculation of Gamma for American put options at different spot prices by varying $\epsilon$. We can see that, by increasing the range of spot prices used to train the interpolator, the noise decreases without generating biases.

\begin{figure}
\ifx\RemoveFigures\undefined
  \begin{center}
  \scalebox{0.95}{%
    \begin{tikzpicture}
      \begin{axis}[set layers, mark layer=axis background,
                  xlabel=Spot Price,
                  ylabel=Delta,
                  ylabel style={overlay},
                  xmin=0.55, xmax=1.45,
                  xtick={0.6,0.8,1,1.2,1.4},
                  xticklabels={60,80,100,120,140},
                  ymin=-1.05, ymax=0.05,
                  grid=major,
                  legend style={legend pos=north west},
                  axis background/.style={fill=gray!10}]
      \addplot [color=mydarkblue!39,thick,smooth] table [y=c32d,x=money] from \senscmp;
      \addplot [color=mydarkblue!69,thick,smooth] table [y=c16d,x=money] from \senscmp;
      \addplot [color=mydarkblue!99,thick,smooth] table [y=c8d,x=money] from \senscmp;
      % \addplot [color=mydarkblue!99,thick,smooth] table [y=c4d,x=money] from \senscmp;
      % \addplot [color=mydarkblue!59,thick,smooth] table [y=ctd,x=money] from \senscmp;
      % \addplot [color=mydarkblue!79,thick,smooth] table [y=cmd,x=money] from \senscmp;
      % \addplot [color=mydarkblue!99,thick,smooth] table [y=ccd,x=money] from \senscmp;
      \addplot [color=mydarkorange,thick,only marks,mark options={scale=.5}] table [y=btd,x=money] from \senscmp;
      \legend{{\small $\epsilon=\sfrac{1}{32}$},{\small $\epsilon=\sfrac{1}{16}$},{\small $\epsilon=1/8$},{\small BT}}
      \end{axis}
    \end{tikzpicture}
    ~~~
    \begin{tikzpicture}
      \begin{axis}[set layers, mark layer=axis background,
                  xlabel=Spot Price,
                  ylabel=Gamma,
                  ylabel style={overlay},
                  xmin=0.55, xmax=1.45,
                  xtick={0.6,0.8,1,1.2,1.4},
                  xticklabels={60,80,100,120,140},
                  ymin=-0.005, ymax=0.035,
                  grid=major,
                  legend style={legend pos=north west},
                  axis background/.style={fill=gray!10}]
      \addplot [color=mydarkblue!39,thick,smooth,select coords between index={1}{21}] table [y=c32g,x=money] from \senscmp;
      \addplot [color=mydarkblue!69,thick,smooth,select coords between index={1}{21}] table [y=c16g,x=money] from \senscmp;
      \addplot [color=mydarkblue!99,thick,smooth,select coords between index={1}{21}] table [y=c8g,x=money] from \senscmp;
      \addplot [color=mydarkblue!39,thick,smooth,select coords between index={21}{99}] table [y=c32g,x=money] from \senscmp;
      \addplot [color=mydarkblue!69,thick,smooth,select coords between index={21}{99}] table [y=c16g,x=money] from \senscmp;
      \addplot [color=mydarkblue!99,thick,smooth,select coords between index={21}{99}] table [y=c8g,x=money] from \senscmp;
      % \addplot [color=mydarkblue!99,thick,smooth] table [y=c4g,x=money] from \senscmp;
      % \addplot [color=mydarkblue!59,thick,smooth] table [y=ctg,x=money] from \senscmp;
      % \addplot [color=mydarkblue!79,thick,smooth] table [y=cmg,x=money] from \senscmp;
      % \addplot [color=mydarkblue!99,thick,smooth] table [y=ccg,x=money] from \senscmp;
      \addplot [color=mydarkorange,thick,only marks,mark options={scale=.5}] table [y=btg,x=money] from \senscmp;
      \end{axis}
    \end{tikzpicture}}%
  \end{center}
\fi
\caption[Delta and Gamma risk matrices for American put options -- Chebyshev]{Delta and Gamma risk matrices for American put options with maturity $T=0.2$, strike $K=100$ and time steps $N=50$. Orange dots are calculated with a standard binomial tree implementation, while blue lines with the Chebyshev method.}
\label{fig:american_put_sensitivity_cheb}
\end{figure}

\subsection{Numerical investigations on fixed-strike Asian sensitivities}
\label{app:sensitivity-asian}

We continue the analysis on sensitivity computations by going beyond the simple, but relevant, case of American options. We find that the Chebyshev method seems a good candidate as target method. Anyway, we prefer to check again our findings also in the case of other payoffs.

In the case of fixed-strike asian payoffs we can see in Figure~\ref{fig:asian_fixed_sensitivity_poly} the behavior of Gamma sensitivity for a moving window of length $M=2$ with different choices of $\epsilon$ when using polynomials for the regression. We obtain the same pattern depicted in the case of American put options. The finite-difference method stabilizes as we increase $\epsilon$, but a visible bias arises, as we can see by comparing the two panels of the Figure.

\begin{figure}
\ifx\RemoveFigures\undefined
  \begin{center}
  \scalebox{0.95}{%
    \begin{tikzpicture}
      \begin{axis}[set layers, mark layer=axis background,
                  xlabel=Spot Price,
                  ylabel=Gamma,
                  ylabel style={overlay},
                  xmin=0.55, xmax=1.45,
                  xtick={0.6,0.8,1,1.2,1.4},
                  xticklabels={60,80,100,120,140},
                  ymin=-0.005, ymax=0.035,
                  grid=major,
                  legend style={legend pos=north east},
                  axis background/.style={fill=gray!10}]
        \addplot [color=mydarkblue!39,thick,smooth] table [y=yge00625,x=money] from \sensaxscmp;
        \addplot [color=mydarkblue!69,thick,smooth] table [y=yge0125,x=money] from \sensaxscmp;
        \addplot [color=mydarkblue!99,thick,smooth] table [y=yge025,x=money] from \sensaxscmp;
        \legend{{\small $\epsilon=\sfrac{1}{16}$},{\small $\epsilon=1/8$},{\small $\epsilon=1/4$}}
    \end{axis}
    \end{tikzpicture}
    ~~~
    \begin{tikzpicture}
      \begin{axis}[set layers, mark layer=axis background,
                  xlabel=Spot Price,
                  ylabel=Gamma,
                  ylabel style={overlay},
                  xmin=0.55, xmax=1.45,
                  xtick={0.6,0.8,1,1.2,1.4},
                  xticklabels={60,80,100,120,140},
                  ymin=-0.005, ymax=0.035,
                  grid=major,
                  legend style={legend pos=north east},
                  axis background/.style={fill=gray!10}]
        \addplot [color=mydarkblue!39,thick,smooth] table [y=ygc025,x=money] from \sensaxscmp;
        \addplot [color=mydarkblue!69,thick,smooth] table [y=ygc0375,x=money] from \sensaxscmp;
        \addplot [color=mydarkblue!99,thick,smooth] table [y=ygc0625,x=money] from \sensaxscmp;
        \legend{{\small $\epsilon=1/4$},{\small $\epsilon=3/8$},{\small $\epsilon=5/8$}}
      \end{axis}
    \end{tikzpicture}}%
  \end{center}
\fi
\caption[Gamma risk matrices for fixed-strike Asian options -- polynomials]{Gamma risk matrices for fixed-strike Asian options with moving window of $M=2$ days, maturity $T=0.2$, strike $K=100$ and time steps $N=50$. LSMC method with polynomial basis functions. Left panel: finite-difference method with different choices of $\epsilon$. Right panel: Chebyshev method with different choices of $\epsilon$.}
\label{fig:asian_fixed_sensitivity_poly}
\end{figure}

For the same fixed-strike Asian payoff we repeat the exercise with R-FFNNs as basis functions. We obtain a very similar result, and we do not notice an increase in noise.

\begin{figure}
\ifx\RemoveFigures\undefined
  \begin{center}
  \scalebox{0.95}{%
    \begin{tikzpicture}
      \begin{axis}[set layers, mark layer=axis background,
                  xlabel=Spot Price,
                  ylabel=Gamma,
                  ylabel style={overlay},
                  xmin=0.55, xmax=1.45,
                  xtick={0.6,0.8,1,1.2,1.4},
                  xticklabels={60,80,100,120,140},
                  ymin=-0.005, ymax=0.035,
                  grid=major,
                  legend style={legend pos=north east},
                  axis background/.style={fill=gray!10}]
        \addplot [color=mydarkblue!39,thick,smooth] table [y=fge00625,x=money] from \sensaxscmp;
        \addplot [color=mydarkblue!69,thick,smooth] table [y=fge0125,x=money] from \sensaxscmp;
        \addplot [color=mydarkblue!99,thick,smooth] table [y=fge025,x=money] from \sensaxscmp;
        \legend{{\small $\epsilon=\sfrac{1}{16}$},{\small $\epsilon=1/8$},{\small $\epsilon=1/4$}}
      \end{axis}
    \end{tikzpicture}
    ~~~
    \begin{tikzpicture}
      \begin{axis}[set layers, mark layer=axis background,
                  xlabel=Spot Price,
                  ylabel=Gamma,
                  ylabel style={overlay},
                  xmin=0.55, xmax=1.45,
                  xtick={0.6,0.8,1,1.2,1.4},
                  xticklabels={60,80,100,120,140},
                  ymin=-0.005, ymax=0.035,
                  grid=major,
                  legend style={legend pos=north east},
                  axis background/.style={fill=gray!10}]
        \addplot [color=mydarkblue!39,thick,smooth] table [y=fgc025,x=money] from \sensaxscmp;
        \addplot [color=mydarkblue!69,thick,smooth] table [y=fgc0375,x=money] from \sensaxscmp;
        \addplot [color=mydarkblue!99,thick,smooth] table [y=fgc0625,x=money] from \sensaxscmp;
        \legend{{\small $\epsilon=1/4$},{\small $\epsilon=3/8$},{\small $\epsilon=5/8$}}
      \end{axis}
    \end{tikzpicture}}%
  \end{center}
\fi
\caption[Gamma risk matrices for fixed-strike Asian options -- R-FFNN]{Gamma risk matrices for fixed-strike Asian options with moving window of $M=2$ days, maturity $T=0.2$, strike $K=100$ and time steps $N=50$. LSMC method with R-FFNN basis functions. Left panel: finite-difference method with different choices of $\epsilon$. Right panel: Chebyshev method with different choices of $\epsilon$.}
\label{fig:asian_fixed_sensitivity_ffnn}
\end{figure}
        
\subsection{Numerical investigations on certificate sensitivities}
\label{app:sensitivity-certificate}

We continue the analysis on sensitivity computations with the certificate case. We start with the snowball version, whose characteristics are described in Section~\ref{subsec:snowball_num}. We show the results for different regression basis functions in Figure~\ref{fig:snowball_sensitivity_poly_ffnn} and~\ref{fig:snowball_sensitivity_rnn_sig}. Polynomials basis function are taken up to third order; random neural networks have $120$ inner nodes; signature methods are truncated at third degree with lead-lag and time embedding. Again we can see that the Chebyshev method, although a little noisier than before, does not show the bias of the finite difference method, which is smoother, but we can see from the diagrams that it moves away from the other curves as soon as the shift $\epsilon$ increases.

\begin{figure}
\ifx\RemoveFigures\undefined
  \begin{center}
  \scalebox{0.95}{%
    \begin{tikzpicture}
      \begin{axis}[set layers, mark layer=axis background,
                  xlabel=Spot Price,
                  ylabel=Gamma,
                  ylabel style={overlay},
                  xmin=0.55, xmax=1.45,
                  xtick={0.6,0.8,1,1.2,1.4},
                  xticklabels={60,80,100,120,140},
                  ymin=-0.00021, ymax=0.00001,
                  grid=major,
                  legend style={legend pos=south east},
                  axis background/.style={fill=gray!10}]
        \addplot [color=mydarkred!39,thick,smooth] table [y=yge0125,x=money] from \senssnwcmp;
        \addplot [color=mydarkred!69,thick,smooth] table [y=yge025,x=money] from \senssnwcmp;
        \addplot [color=mydarkred!99,thick,smooth] table [y=yge0375,x=money] from \senssnwcmp;
        \addplot [color=mydarkblue!39,thick,smooth] table [y=ygc05,x=money] from \senssnwcmp;
        \addplot [color=mydarkblue!69,thick,smooth] table [y=ygc0625,x=money] from \senssnwcmp;
        \addplot [color=mydarkblue!99,thick,smooth] table [y=ygc075,x=money] from \senssnwcmp;
        \legend{{\small FD $\epsilon=\sfrac{1}{8}$},{\small FD $\epsilon=\sfrac{1}{4}$},{\small FD $\epsilon=\sfrac{3}{8}$},{\small Cheb $\epsilon=\sfrac{1}{2}$},{\small Cheb $\epsilon=\sfrac{5}{8}$},{\small Cheb $\epsilon=\sfrac{3}{4}$}}
      \end{axis}
    \end{tikzpicture}
    ~~~
    \begin{tikzpicture}
      \begin{axis}[set layers, mark layer=axis background,
                  xlabel=Spot Price,
                  ylabel=Gamma,
                  ylabel style={overlay},
                  xmin=0.55, xmax=1.45,
                  xtick={0.6,0.8,1,1.2,1.4},
                  xticklabels={60,80,100,120,140},
                  ymin=-0.00021, ymax=0.00001,
                  grid=major,
                  legend style={legend pos=south east},
                  axis background/.style={fill=gray!10}]
        \addplot [color=mydarkred!39,thick,smooth] table [y=fge0125,x=money] from \senssnwcmp;
        \addplot [color=mydarkred!69,thick,smooth] table [y=fge025,x=money] from \senssnwcmp;
        \addplot [color=mydarkred!99,thick,smooth] table [y=fge0375,x=money] from \senssnwcmp;
        \addplot [color=mydarkblue!39,thick,smooth] table [y=fgc05,x=money] from \senssnwcmp;
        \addplot [color=mydarkblue!69,thick,smooth] table [y=fgc0625,x=money] from \senssnwcmp;
        \addplot [color=mydarkblue!99,thick,smooth] table [y=fgc075,x=money] from \senssnwcmp;
        \legend{{\small FD $\epsilon=\sfrac{1}{8}$},{\small FD $\epsilon=\sfrac{1}{4}$},{\small FD $\epsilon=\sfrac{3}{8}$},{\small Cheb $\epsilon=\sfrac{1}{2}$},{\small Cheb $\epsilon=\sfrac{5}{8}$},{\small Cheb $\epsilon=\sfrac{3}{4}$}}
      \end{axis}
    \end{tikzpicture}}%
  \end{center}
\fi
\caption[Gamma risk matrices for snowball certificates -- polynomials and R-FFNN]{Gamma risk matrices for snowball certificates, maturity $T=2$. LSMC method with polynomial basis functions (left panel), R-FFNN basis functions (right panel). Finite difference method in red, Chebyshev method in blue.}
\label{fig:snowball_sensitivity_poly_ffnn}
\end{figure}

\begin{figure}
\ifx\RemoveFigures\undefined
  \begin{center}
  \scalebox{0.95}{%
    \begin{tikzpicture}
      \begin{axis}[set layers, mark layer=axis background,
                  xlabel=Spot Price,
                  ylabel=Gamma,
                  ylabel style={overlay},
                  xmin=0.55, xmax=1.45,
                  xtick={0.6,0.8,1,1.2,1.4},
                  xticklabels={60,80,100,120,140},
                  ymin=-0.00021, ymax=0.00001,
                  grid=major,
                  legend style={legend pos=south east},
                  axis background/.style={fill=gray!10}]
        \addplot [color=mydarkred!39,thick,smooth] table [y=rge0125,x=money] from \senssnwcmp;
        \addplot [color=mydarkred!69,thick,smooth] table [y=rge025,x=money] from \senssnwcmp;
        \addplot [color=mydarkred!99,thick,smooth] table [y=rge0375,x=money] from \senssnwcmp;
        \addplot [color=mydarkblue!39,thick,smooth] table [y=rgc05,x=money] from \senssnwcmp;
        \addplot [color=mydarkblue!69,thick,smooth] table [y=rgc0625,x=money] from \senssnwcmp;
        \addplot [color=mydarkblue!99,thick,smooth] table [y=rgc075,x=money] from \senssnwcmp;
        \legend{{\small FD $\epsilon=\sfrac{1}{8}$},{\small FD $\epsilon=\sfrac{1}{4}$},{\small FD $\epsilon=\sfrac{3}{8}$},{\small Cheb $\epsilon=\sfrac{1}{2}$},{\small Cheb $\epsilon=\sfrac{5}{8}$},{\small Cheb $\epsilon=\sfrac{3}{4}$}}
      \end{axis}
    \end{tikzpicture}
    ~~~
    \begin{tikzpicture}
      \begin{axis}[set layers, mark layer=axis background,
                  xlabel=Spot Price,
                  ylabel=Gamma,
                  ylabel style={overlay},
                  xmin=0.55, xmax=1.45,
                  xtick={0.6,0.8,1,1.2,1.4},
                  xticklabels={60,80,100,120,140},
                  ymin=-0.00021, ymax=0.00001,
                  grid=major,
                  legend style={legend pos=south east},
                  axis background/.style={fill=gray!10}]
        \addplot [color=mydarkred!39,thick,smooth] table [y=sge0125,x=money] from \senssnwcmp;
        \addplot [color=mydarkred!69,thick,smooth] table [y=sge025,x=money] from \senssnwcmp;
        \addplot [color=mydarkred!99,thick,smooth] table [y=sge0375,x=money] from \senssnwcmp;
        \addplot [color=mydarkblue!39,thick,smooth] table [y=sgc05,x=money] from \senssnwcmp;
        \addplot [color=mydarkblue!69,thick,smooth] table [y=sgc0625,x=money] from \senssnwcmp;
        \addplot [color=mydarkblue!99,thick,smooth] table [y=sgc075,x=money] from \senssnwcmp;
        \legend{{\small FD $\epsilon=\sfrac{1}{8}$},{\small FD $\epsilon=\sfrac{1}{4}$},{\small FD $\epsilon=\sfrac{3}{8}$},{\small Cheb $\epsilon=\sfrac{1}{2}$},{\small Cheb $\epsilon=\sfrac{5}{8}$},{\small Cheb $\epsilon=\sfrac{3}{4}$}}
      \end{axis}
    \end{tikzpicture}}%
  \end{center}
\fi
\caption[Gamma risk matrices for snowball certificates -- R-RNN and signatures]{Gamma risk matrices for snowball certificates, maturity $T=2$. LSMC method with R-RNN basis functions (left panel), signature basis functions (right panel). Finite difference method red, Chebyshev method in blue.}
\label{fig:snowball_sensitivity_rnn_sig}
\end{figure}

We continue with the lock-in version, whose characteristics are described in Section~\ref{subsec:lockin_num}. We show the results for different regression basis functions in Figure~\ref{fig:lockin_sensitivity_poly_ffnn} and~\ref{fig:lockin_sensitivity_rnn_sig}. Polynomials basis function are taken up to third order; random neural networks have $120$ inner nodes; signature methods are truncated at third degree with lead-lag and time embedding. We can see that the finite-difference method is quite biased, while the Chebyshev method seems working well.

\begin{figure}
\ifx\RemoveFigures\undefined
  \begin{center}
  \scalebox{0.95}{%
    \begin{tikzpicture}
      \begin{axis}[set layers, mark layer=axis background,
                  xlabel=Spot Price,
                  ylabel=Gamma,
                  ylabel style={overlay},
                  xmin=0.55, xmax=1.45,
                  xtick={0.6,0.8,1,1.2,1.4},
                  xticklabels={60,80,100,120,140},
                  ymin=-0.00021, ymax=0.00006,
                  grid=major,
                  legend style={legend pos=south east},
                  axis background/.style={fill=gray!10}]
        \addplot [color=mydarkred!39,thick,smooth] table [y=yge0125,x=money] from \senslckcmp;
        \addplot [color=mydarkred!69,thick,smooth] table [y=yge025,x=money] from \senslckcmp;
        \addplot [color=mydarkred!99,thick,smooth] table [y=yge0375,x=money] from \senslckcmp;
        \addplot [color=mydarkblue!39,thick,smooth] table [y=ygc05,x=money] from \senslckcmp;
        \addplot [color=mydarkblue!69,thick,smooth] table [y=ygc0625,x=money] from \senslckcmp;
        \addplot [color=mydarkblue!99,thick,smooth] table [y=ygc075,x=money] from \senslckcmp;
        \legend{{\small FD $\epsilon=\sfrac{1}{8}$},{\small FD $\epsilon=\sfrac{1}{4}$},{\small FD $\epsilon=\sfrac{3}{8}$},{\small Cheb $\epsilon=\sfrac{1}{2}$},{\small Cheb $\epsilon=\sfrac{5}{8}$},{\small Cheb $\epsilon=\sfrac{3}{4}$}}
      \end{axis}
    \end{tikzpicture}
    ~~~
    \begin{tikzpicture}
      \begin{axis}[set layers, mark layer=axis background,
                  xlabel=Spot Price,
                  ylabel=Gamma,
                  ylabel style={overlay},
                  xmin=0.55, xmax=1.45,
                  xtick={0.6,0.8,1,1.2,1.4},
                  xticklabels={60,80,100,120,140},
                  ymin=-0.00021, ymax=0.00006,
                  grid=major,
                  legend style={legend pos=south east},
                  axis background/.style={fill=gray!10}]
        \addplot [color=mydarkred!39,thick,smooth] table [y=fge0125,x=money] from \senslckcmp;
        \addplot [color=mydarkred!69,thick,smooth] table [y=fge025,x=money] from \senslckcmp;
        \addplot [color=mydarkred!99,thick,smooth] table [y=fge0375,x=money] from \senslckcmp;
        \addplot [color=mydarkblue!39,thick,smooth] table [y=fgc05,x=money] from \senslckcmp;
        \addplot [color=mydarkblue!69,thick,smooth] table [y=fgc0625,x=money] from \senslckcmp;
        \addplot [color=mydarkblue!99,thick,smooth] table [y=fgc075,x=money] from \senslckcmp;
        \legend{{\small FD $\epsilon=\sfrac{1}{8}$},{\small FD $\epsilon=\sfrac{1}{4}$},{\small FD $\epsilon=\sfrac{3}{8}$},{\small Cheb $\epsilon=\sfrac{1}{2}$},{\small Cheb $\epsilon=\sfrac{5}{8}$},{\small Cheb $\epsilon=\sfrac{3}{4}$}}
      \end{axis}
    \end{tikzpicture}}%
  \end{center}
\fi
\caption[Gamma risk matrices for lock-in certificates -- polynomials and R-FFNN]{Gamma risk matrices for lock-in certificates, maturity $T=2$. LSMC method with polynomial basis functions (left panel), R-FFNN basis functions (right panel). Finite difference method in red, Chebyshev method in blue.}
\label{fig:lockin_sensitivity_poly_ffnn}
\end{figure}

\begin{figure}
\ifx\RemoveFigures\undefined
  \begin{center}
  \scalebox{0.95}{%
    \begin{tikzpicture}
      \begin{axis}[set layers, mark layer=axis background,
                  xlabel=Spot Price,
                  ylabel=Gamma,
                  ylabel style={overlay},
                  xmin=0.55, xmax=1.45,
                  xtick={0.6,0.8,1,1.2,1.4},
                  xticklabels={60,80,100,120,140},
                  ymin=-0.00021, ymax=0.00006,
                  grid=major,
                  legend style={legend pos=south east},
                  axis background/.style={fill=gray!10}]
        \addplot [color=mydarkred!39,thick,smooth] table [y=rge0125,x=money] from \senslckcmp;
        \addplot [color=mydarkred!69,thick,smooth] table [y=rge025,x=money] from \senslckcmp;
        \addplot [color=mydarkred!99,thick,smooth] table [y=rge0375,x=money] from \senslckcmp;
        \addplot [color=mydarkblue!39,thick,smooth] table [y=rgc05,x=money] from \senslckcmp;
        \addplot [color=mydarkblue!69,thick,smooth] table [y=rgc0625,x=money] from \senslckcmp;
        \addplot [color=mydarkblue!99,thick,smooth] table [y=rgc075,x=money] from \senslckcmp;
        \legend{{\small FD $\epsilon=\sfrac{1}{8}$},{\small FD $\epsilon=\sfrac{1}{4}$},{\small FD $\epsilon=\sfrac{3}{8}$},{\small Cheb $\epsilon=\sfrac{1}{2}$},{\small Cheb $\epsilon=\sfrac{5}{8}$},{\small Cheb $\epsilon=\sfrac{3}{4}$}}
      \end{axis}
    \end{tikzpicture}
    ~~~
    \begin{tikzpicture}
      \begin{axis}[set layers, mark layer=axis background,
                  xlabel=Spot Price,
                  ylabel=Gamma,
                  ylabel style={overlay},
                  xmin=0.55, xmax=1.45,
                  xtick={0.6,0.8,1,1.2,1.4},
                  xticklabels={60,80,100,120,140},
                  ymin=-0.00021, ymax=0.00006,
                  grid=major,
                  legend style={legend pos=south east},
                 axis background/.style={fill=gray!10}]
        \addplot [color=mydarkred!39,thick,smooth] table [y=sge0125,x=money] from \senslckcmp;
        \addplot [color=mydarkred!69,thick,smooth] table [y=sge025,x=money] from \senslckcmp;
        \addplot [color=mydarkred!99,thick,smooth] table [y=sge0375,x=money] from \senslckcmp;
        \addplot [color=mydarkblue!39,thick,smooth] table [y=sgc05,x=money] from \senslckcmp;
        \addplot [color=mydarkblue!69,thick,smooth] table [y=sgc0625,x=money] from \senslckcmp;
        \addplot [color=mydarkblue!99,thick,smooth] table [y=sgc075,x=money] from \senslckcmp;
        \legend{{\small FD $\epsilon=\sfrac{1}{8}$},{\small FD $\epsilon=\sfrac{1}{4}$},{\small FD $\epsilon=\sfrac{3}{8}$},{\small Cheb $\epsilon=\sfrac{1}{2}$},{\small Cheb $\epsilon=\sfrac{5}{8}$},{\small Cheb $\epsilon=\sfrac{3}{4}$}}
      \end{axis}
    \end{tikzpicture}}%
  \end{center}
\fi
\caption[Gamma risk matrices for lock-in certificates -- R-RNN and signatures]{Gamma risk matrices for lock-in certificates, maturity $T=2$. LSMC method with R-RNN basis functions (left panel), signature basis functions (right panel). Finite difference method red, Chebyshev method in blue.}
\label{fig:lockin_sensitivity_rnn_sig}
\end{figure}

\section{Supplemental material}
\label{app:numerical_data}

We collect in this section further numerical results, which support the discussion of the main body of the paper. In Tables~\ref{tab:afs_errors}, \ref{tab:axs_errors}, \ref{tab:lfs_errors}, \ref{tab:lxs_errors} and~\ref{tab:certificate_errors} we report the statistical uncertainties of Monte Carlo simulations used to price all the contracts discussed in the main body of the paper. Statistical uncertainties are reported at one-sigma confidence level.

\begin{table}
 \begin{center}
  \pgfplotstabletypeset[
    print last=35,
    skip rows between index={29}{31},
    string replace={0}{},
    fixed, fixed zerofill, precision=3,
    every head row/.style={before row=\toprule\multicolumn{8}{l}{\bf Floating-strike Asian option: statistical uncertainties}\\\toprule,after row=\midrule},
    every last row/.style={after row=\bottomrule},
    every row no 4/.style={before row=\hline},
    every row no 8/.style={before row=\hline},
    every row no 12/.style={before row=\hline},
    every row no 16/.style={before row=\hline},
    every row no 20/.style={before row=\hline},
    every row no 24/.style={before row=\hline},
    every row no 27/.style={before row=\hline},
    columns={model,afs2e,afs3e,afs4e,afs5e,afs10e,afs20e,afs30e},
    columns/model/.style={string type,column type=l|,column name={\bf Model}},
    columns/afs2e/.style={column name={$\mathbf{2}$}},
    columns/afs3e/.style={column name={$\mathbf{3}$}},
    columns/afs4e/.style={column name={$\mathbf{4}$}},
    columns/afs5e/.style={column name={$\mathbf{5}$}},
    columns/afs10e/.style={column name={$\mathbf{10}$}},
    columns/afs20e/.style={column name={$\mathbf{20}$}},
    columns/afs30e/.style={column name={$\mathbf{30}$}},
  ]{\afsdat}
 \end{center}
 \caption[Statistical uncertainties of floating-strike American-style Asian options]{Statistical uncertainties of the Monte Carlo simulation used to price floating-strike American-style Asian options with strike $K=100$, maturity $T=0.2$ and time steps $N=50$ for different lengths of the moving window. Comparison between different models (see text). Benchmark models: GPR-GHQ and binomial Markov chain of~\textcite{Goudenege2022}, \textcite{bernhart2011finite}, \textcite{lelong2019pricing}. LSMC models: polynomial bases, randomized neural networks (R-FFNN and R-RNN), signature and randomized signature (RandSig) based methods.}
 \label{tab:afs_errors}
\end{table}

\begin{table}
 \begin{center}
  \pgfplotstabletypeset[
    skip rows between index={24}{26},
    string replace={0}{},
    fixed, fixed zerofill, precision=3,
    every head row/.style={before row=\toprule\multicolumn{8}{l}{\bf Fixed-strike Asian option: statistical uncertainties}\\\toprule,after row=\midrule},
    every last row/.style={after row=\bottomrule},
    every row no 4/.style={before row=\hline},
    every row no 8/.style={before row=\hline},
    every row no 12/.style={before row=\hline},
    every row no 16/.style={before row=\hline},
    every row no 20/.style={before row=\hline},
    every row no 24/.style={before row=\hline},
    every row no 28/.style={before row=\hline},
    columns={model,axs2e,axs3e,axs4e,axs5e,axs10e,axs20e,axs30e},
    columns/model/.style={string type,column type=l|,column name={\bf Model}},
    columns/axs2e/.style={column name={$\mathbf{2}$}},
    columns/axs3e/.style={column name={$\mathbf{3}$}},
    columns/axs4e/.style={column name={$\mathbf{4}$}},
    columns/axs5e/.style={column name={$\mathbf{5}$}},
    columns/axs10e/.style={column name={$\mathbf{10}$}},
    columns/axs20e/.style={column name={$\mathbf{20}$}},
    columns/axs30e/.style={column name={$\mathbf{30}$}},
  ]{\axsdat}
 \end{center}
 \caption[Statistical uncertainties of fixed-strike American-style Asian options]{Statistical uncertainties of the Monte Carlo simulation used to price fixed-strike American-style Asian options with strike $K=100$, maturity $T=0.2$ and time steps $N=50$ for different lengths of the moving window. Comparison between different models (see text). LSMC models: polynomial bases, randomized neural networks (R-FFNN and R-RNN), signature and randomized signature (RandSig) based methods.}
 \label{tab:axs_errors}
\end{table}

\begin{table}
 \begin{center}
  \pgfplotstabletypeset[
    skip rows between index={24}{26},
    string replace={0}{},
    fixed, fixed zerofill, precision=3,
    every head row/.style={before row=\toprule\multicolumn{8}{l}{\bf Floating-strike look-back option: statistical uncertainties}\\\toprule,after row=\midrule},
    every last row/.style={after row=\bottomrule},
    every row no 4/.style={before row=\hline},
    every row no 8/.style={before row=\hline},
    every row no 12/.style={before row=\hline},
    every row no 16/.style={before row=\hline},
    every row no 20/.style={before row=\hline},
    every row no 24/.style={before row=\hline},
    columns={model,lfs2e,lfs3e,lfs4e,lfs5e,lfs10e,lfs20e,lfs30e},
    columns/model/.style={string type,column type=l|,column name={\bf Model}},
    columns/lfs2e/.style={column name={$\mathbf{2}$}},
    columns/lfs3e/.style={column name={$\mathbf{3}$}},
    columns/lfs4e/.style={column name={$\mathbf{4}$}},
    columns/lfs5e/.style={column name={$\mathbf{5}$}},
    columns/lfs10e/.style={column name={$\mathbf{10}$}},
    columns/lfs20e/.style={column name={$\mathbf{20}$}},
    columns/lfs30e/.style={column name={$\mathbf{30}$}},
  ]{\lfsdat}
 \end{center}
 \caption[Statistical uncertainties of floating-strike American-style look-back options]{Statistical uncertainties of the Monte Carlo simulation used to price floating-strike American-style look-back options with maturity $T=0.2$ and time steps $N=50$ for different lengths of the moving window. Comparison between different models (see text). LSMC models: polynomial bases, randomized neural networks (R-FFNN and R-RNN), signature based methods.}
 \label{tab:lfs_errors}
\end{table}

\begin{table}
 \begin{center}
  \pgfplotstabletypeset[
    skip rows between index={24}{26},
    string replace={0}{},
    fixed, fixed zerofill, precision=3,
    every head row/.style={before row=\toprule\multicolumn{8}{l}{\bf Fixed-strike look-back option: statistical uncertainties}\\\toprule,after row=\midrule},
    every last row/.style={after row=\bottomrule},
    every row no 4/.style={before row=\hline},
    every row no 8/.style={before row=\hline},
    every row no 12/.style={before row=\hline},
    every row no 16/.style={before row=\hline},
    every row no 20/.style={before row=\hline},
    every row no 24/.style={before row=\hline},
    columns={model,lxs2e,lxs3e,lxs4e,lxs5e,lxs10e,lxs20e,lxs30e},
    columns/model/.style={string type,column type=l|,column name={\bf Model}},
    columns/lxs2e/.style={column name={$\mathbf{2}$}},
    columns/lxs3e/.style={column name={$\mathbf{3}$}},
    columns/lxs4e/.style={column name={$\mathbf{4}$}},
    columns/lxs5e/.style={column name={$\mathbf{5}$}},
    columns/lxs10e/.style={column name={$\mathbf{10}$}},
    columns/lxs20e/.style={column name={$\mathbf{20}$}},
    columns/lxs30e/.style={column name={$\mathbf{30}$}},
  ]{\lxsdat}
 \end{center}
 \caption[Statistical uncertainties of fixed-strike American-style look-back options]{Statistical uncertainties of the Monte Carlo simulation used to price fixed-strike American-style look-back options with strike $K=100$, maturity $T=0.2$ and time steps $N=50$ for different lengths of the moving window. Comparison between different models (see text). LSMC models: polynomial bases, randomized neural networks (R-FFNN and R-RNN), signature based methods.}
 \label{tab:lxs_errors}
\end{table}

\begin{table}
 \begin{center}
  \pgfplotstabletypeset[
    string replace={0}{},
    fixed, fixed zerofill, precision=5,
    every head row/.style={before row=\toprule{\bf Certificate: uncert.}&\multicolumn{3}{c|}{\bf Snowball}&\multicolumn{3}{c}{\bf Lock-in}\\\toprule,after row=\midrule},
    every last row/.style={after row=\bottomrule},
    every row no 3/.style={before row=\hline},
    every row no 6/.style={before row=\hline},
    every row no 9/.style={before row=\hline},
    columns={model,sn1e,sn2e,sn5e,li1e,li2e,li5e},
    columns/model/.style={string type,column type=l|,column name={\bf Model}},
    columns/sn1e/.style={column name={\bf 1y}},
    columns/sn2e/.style={column name={\bf 2y}},
    columns/sn5e/.style={column type=c|,column name={\bf 5y}},
    columns/li1e/.style={column name={\bf 1y}},
    columns/li2e/.style={column name={\bf 2y}},
    columns/li5e/.style={column name={\bf 5y}},
  ]{\certificatedat}
 \end{center}
 \caption[Statistical uncertainties of snowball and lock-in certificates]{Statistical uncertainties of the Monte Carlo simulation used to price snowball and lock-in certificates with maturity of 1, 2 and 5 years (other characteristics described in the text). Comparison between different models (see text). LSMC models: polynomial bases, randomized neural networks (R-FFNN and R-RNN), signature based methods.}
 \label{tab:certificate_errors}
\end{table}

\end{document}